\documentclass[12pt]{article}
\usepackage{amsmath}
\usepackage{amssymb}
\usepackage{amstext}
\usepackage{amsfonts}
\usepackage{graphicx,epsfig}
\usepackage{slashbox}
\usepackage{indentfirst}

\def \bea{\begin{eqnarray}}
\def \beq{\begin{equation}}

\def \eea{\end{eqnarray}}
\def \eeq{\end{equation}}

\begin{document}

\baselineskip 20pt

\title{Baryonium Study in Heavy Baryon Chiral Perturbation Theory}
\author{Yue-De Chen$^a$\footnote{Email: chenyuede-b07@mails.gucas.ac.cn}\;
and
Cong-Feng Qiao$^{a,b}\footnote{Email: qiaocf@gucas.ac.cn}$\\[0.5cm]
\small $a)$ Department of Physics, Graduate University, the Chinese
Academy of Sciences \\ \small YuQuan Road 19A, 100049, Beijing,
China\\
 {\small $b)$ Theoretical Physics Center for Science Facilities
(TPCSF), CAS}\\
{\small YuQuan Road 19B, 100049, Beijing, China}\\
}
\date{}
\maketitle

\begin{center}
\begin{minipage}{11cm}
To see whether heavy baryon and anti-baryon can form a bound state,
the heavy baryonium, we study the two-pion exchange interaction
potential between them within the heavy baryon chiral perturbation
theory. The obtained potential is applied to calculate the heavy
baryonium masses by solving the Schr\"{o}dinger equation. We find it
is true that the heavy baryonium may exist in a reasonable choice of
input parameters. The uncertainties remaining in the potential and
their influences on the heavy baryonium mass spectrum are discussed.
\end{minipage}
\end{center}

\section{Introduction}

Quark model has achieved great success in describing the
experimentally observed hadronic structures to a large extent. And
the quark potential in between quark and anti-quark deduced from
Chromodynamics (QCD) can explain the meson spectrum quite well. Many
of predicted states by potential model were discovered in experiment
and the theoretical predictions are in good agreement with
experimental data, especially in charmonium and bottomonium sectors
\cite{w_Lucha, c_quigg, v_novikov}, where the masses of charm and
bottom quarks are heavy enough to be treated non-relativistically.
However, things became confused after the discovery of $X(3872)$ in
2003 at $\mathrm{Belle}$ \cite{belle}, which was later confirmed by
$\mathrm{BaBar}$ \cite{babar}. In recent years, a series of unusual
states in charmonium sector, such as $Y(4260)$, $Y(4360)$,
$Y(4660)$, and $Z^{\pm}(4430)$, were observed in experiment
\cite{ex}. Due to their extraordinary decay nature, it is hard to
embed them into the conventional charmonium spectrum, which leads
people to treat them as exotic rather than quark-quark bound states.
The typical scenarios in explaining these newly found states include
treating $Y(4260)$ as a hybrid charmonium \cite{Y4260}, a
$\chi_{c}\rho^{0}$ molecular state \cite{Y4260-Liu}, a conventional
$\Psi(4S)$ \cite{Y4260-FJ}, an $\omega\chi_{c1}$ molecular state
\cite{Y4260-Yuan}, a $\Lambda_{c}\bar{\Lambda}_{c}$ baryonium state
\cite{Y4260-Qiao}, a $D_{1}D$ or $D_{0}D^{*}$ hadronic molecule
\cite{Y4260-Ding}, and a $P$-wave tetraquark $[cs][\bar{c}\bar{s}]$
state \cite{Y4260-L}; $Y(4360)$ is interpreted as the candidate of
the charmonium hybrid or an excited D-wave charmonium state, the
$3^{3}D_{1}$ \cite{Ding} and an excited state of baryonium
\cite{Qiao}; $Y(4660)$ is suggested to be the excited S-wave
charmonium states, the $5^{3}S_{1}$ \cite{Ding} and $6^{3}S_{1}$
\cite{KTChao}, a baryonium state \cite{Qiao,Y4660-Bugg}, a
$f_{0}(980)\Psi'$ bound state \cite{Guo,zgwang}, a
$5^{3}S_{1}$-$4^{3}D_{1}$ mixing state \cite{Badalian}, and also a
tetraquark state \cite{Y4660-QCDSR, Y4660-Ebert-2}. There have been
recently many research works on "exotic" heavy quarkonium study in
experiment and theory. To know more of recent progress in this
respect and to have a more complete list of references one can see
e.g. recent reviews \cite{N.Bram,N.Der} and references therein.

In the baryonium picture, the tri-quark clusters are baryon-like,
but not necessarily colorless. In the pioneer works of heavy
baryonium for the interpretation of newly observed ``exotic"
structures \cite{Y4260-Qiao,Qiao}, there were only phenomenological
and kinematic analysis, but without dynamics. In this work we
attempt to study the heavy baryonium interaction potential arising
from two-pion exchanges in the framework of Heavy Baryon Chiral
Perturbation Theory (HBCPT) \cite{T.M.Yan}. The paper is organized
as follows. In Section 2, we present the formalism for the heavy
baryon-baryon interaction study; in Section 3 we perform the
numerical study for the mass spectrum of the possible baryonium with
the obtained potential in preceding section; the Section 4 is
devoted to the summary and conclusions. For the sake of reader's
convenience some of the used formulae are given in the Appendix.

\section{Formalism}

To obtain the heavy baryonium mass spectrum, we first start from
extracting the baryon-baryon interaction potential in the same
procedure as for quark-quark interaction \cite{w_Lucha}.

\subsection{Heavy Baryonium}

In the heavy baryonium picture \cite{Qiao}, $\Lambda_c$ and
$\Sigma^0_c$ are taken as basis vectors in two-dimensional space.
The baryonia are loosely bound states of heavy baryon and
anti-baryon, namely
\begin{eqnarray}
B^+_1&\equiv &|\Lambda_c^+ \; \bar{\Sigma}_c^0>~~~~~~~~~\nonumber\\
{\rm Triplet:}\;\;\;\;\; B^0_1&\equiv &
\frac{1}{\sqrt{2}}(|\Lambda_c^+
\;\bar{\Lambda}_c^+>\; -\; |{\Sigma}_c^0 \bar{\Sigma}_c^0>)\\
B^-_1&\equiv&|\bar{\Lambda}^+_c\; {\Sigma}_c^0>~~~~~~~~~\nonumber
 \label{triplet}
\end{eqnarray}
and
\begin{eqnarray}
{\rm Singlet:}\;\;\;\;\; B^0_0\equiv \frac{1}{\sqrt{2}}(|\Lambda_c^+
\;\bar{\Lambda}_c^+>\; + \; |{\Sigma}_c^0 \bar{\Sigma}_c^0>)\ .
\label{sin}
\end{eqnarray}
Here, approximately the transformation in this two-dimensional
"C-spin" space is invariant, which is in analog to the invariance of
isospin transformation in proton and neutron system.

\subsection{Effective Chiral Lagrangian}

Heavy baryon contains both light and heavy quarks, of which the
light component exhibits the chiral property and the heavy component
exhibits heavy symmetry. Therefore, it is plausible to tackle the
problem of heavy baryon interaction through the heavy chiral
perturbation theory. Following we briefly review the gists of the
HBCPT for later use.

In usual chiral perturbation theory, the nonlinear chiral symmetry
is realized by making use of the unitary matrix
\begin{equation}
\Sigma=e^{\frac{2i M}{f_\pi}}\; ,
\end{equation}
where $M$ is a $3\times3$ matrix composed of eight Goldstone-boson
fields, i.e.,
\beq
 M =
\left(\begin{array}{ccc} \frac{1}{\sqrt{2}} \pi^0 +
\frac{1}{\sqrt{6}} \eta &\phantom{+} \pi^+
& \phantom{+} K^+ \\
\phantom{+}  \pi^- &  -\frac{1}{\sqrt{2}} \pi^0 + \frac{1}{\sqrt{6}}
\eta & \phantom{+} K^0\\
\phantom{+} K^- &  \phantom{+} \bar{K}^0 & \phantom{+} -
\frac{2}{\sqrt{6}} \eta  \end{array}\right) \; . \eeq
Here, $ f_\pi $ is the $pion$ decay constant.

After the chiral symmetry spontaneously broken, the Goldstone boson
interaction with hadron is introduced through a new matrix
\cite{A.Manohar, M.Wise}
\begin{equation}
\xi=\Sigma^{\frac{1}{2}}=e^{\frac{iM}{f_\pi}}\; .
\end{equation}
From $\xi$ one can construct a vector field $V_\mu$ and an axial
vector field $A_\mu$ with simple chiral transformation properties,
i.e.,
\begin{equation}
V_\mu=\frac{1}{2}(\xi^{\dag}\partial_{\mu}\xi+\xi\partial_{\mu}\xi^{\dag})\;
,
\end{equation}
\begin{equation}
A_\mu=\frac{i}{2}(\xi^{\dag}\partial_{\mu}\xi-\xi\partial_{\mu}\xi^{\dag})\;
.
\end{equation}
For our aim, we work only on the leading order vector and axial
vector fields in the expansion of $\xi$ in terms of $f_\pi$, they
are
\begin{equation}
V_\mu=\frac{1}{f_\pi^2}M\partial_\mu M\; , \label{vector-current}
\end{equation}
\begin{equation}
A_\mu=-\frac{1}{f_\pi}\partial_\mu M\; .
\end{equation}

For heavy baryon, each of the two light quarks is in a triplet of
flavor SU(3), and hence the baryons can be grouped in two different
SU(3) multiplets, the sixtet and antitriplet. The symmetric sixtet
and antisymmetric triplet can be constructed out in $3\times 3$
matrices \cite{M.Wise}, they are
\begin{equation}
B_6=\left(\begin{array}{ccc}\Sigma_c^{++}& \frac{1}
{\sqrt{2}}\Sigma_c^{+} & \frac{1}{\sqrt{2}}\Xi_c^{'+}\\
\frac{1}{\sqrt{2}}\Sigma_c^{+} & \Sigma_c^0 &
\frac{1}{\sqrt{2}}\Xi_c^{'0}\\
\frac{1}{\sqrt{2}}\Xi_c^{'+} & \frac{1}{\sqrt{2}}\Xi_c^{'0} &
\Omega_c^0\end{array}\right)\; ,
\end{equation}
and
\begin{equation}
B_{\bar{3}}=\left(\begin{array}{ccc}0& \Lambda_c & \Xi_c^{+}\\
-\Lambda_c & 0 &
\Xi_c^{-}\\
-\Xi_c^{+} & -\Xi_c^{-} & 0\end{array}\right)\; ,
\end{equation}
respectively.

Introducing six coupling constant $g_i$, $i=1,6$, the general
chiral-invariant Lagrangian then reads \cite{T.M.Yan}
\begin{eqnarray}
\mathcal{L_G}& =
&\frac{1}{2}tr[\bar{B}_{\bar{3}}(iD\!\!\!/-M_{\bar{3}})
B_{\bar{3}}]+tr[\bar{B}_6(iD\!\!\!/-M_6)B_6]\nonumber\\
&+&tr[\bar{B}_6^{*\mu}[-g_{\mu\nu}(iD\!\!\!/-M_6^{*})+i(\gamma_\mu
D_\nu+\gamma_\nu
D_\mu)-\gamma_\mu(iD\!\!\!/+M_6^{*})\gamma_\nu]B_6^{*\nu}]\nonumber\\
&+&g_1tr(\bar{B}_6\gamma_{\mu}\gamma_5A^{\mu}B_6)+g_2tr(\bar{B}_6
\gamma_{\mu}\gamma_5A^{\mu}B_{\bar{3}})+ h.c.\nonumber\\
&+&g_3tr(\bar{B}_{6{\mu}}^*A^{\mu}B_6)+ h.c. + g_4
tr(\bar{B}_{6{\mu}}^*A^{\mu}B_{\bar{3}}) + h.c. \nonumber\\
&+&g_5tr(\bar{B}_6^{\nu*}\gamma_{\mu}\gamma_5A^{\mu}
B_{6\nu}^*)+g_6tr(\bar{B}_{\bar{3}}\gamma_{\mu}\gamma_5A^{\mu}B_{\bar{3}})\;
. \label{general-lag}
\end{eqnarray}
Here, $B_{6\nu}^*$ is a Rarita-Schwinger vector-spinor field for
spin-$\frac{3}{2}$ particle; $M_{\bar{3}}$, $M_6$, $M_6^*$ represent
for heavy baryon mass matrices of corresponding fields; With the
help of vector current $V_\mu$ defined in
Eq.~(\ref{vector-current}), we may construct the covariant
derivative $D_\mu$, which acts on baryon field, as
\begin{equation}
D_\mu B_6 = \partial_\mu B_6 + V_\mu B_6 + B_6 V_\mu ^T \;,
\end{equation}
\begin{equation}
D_\mu B_{\bar{3}} = \partial_\mu B_{\bar{3}} + V_\mu B_{\bar{3}} +
B_{\bar{3}} V_\mu ^T \;,
\end{equation}
where $V_\mu ^T$ stands for the transpose of $V_\mu$. Thus, the
couplings of vector current to heavy baryons relevant to our task
take the following form
\begin{eqnarray}
\mathcal{L}_{{\mathcal{E}_1}} & = & \frac{1}{2}
tr(\bar{B}_{\bar{3}}i\gamma^\mu V_\mu
B_{\bar{3}})\nonumber\\
& = & \frac{1}{2 f_\pi ^2}\bar{\Lambda}_c i\gamma^\mu (\pi^0
\partial_\mu \pi^0 + \pi^{-}\partial_\mu \pi^{+} + \pi^{+}\partial_\mu
\pi^{-})\Lambda_c \;,
\end{eqnarray}
and
\begin{eqnarray}
\mathcal{L}_{{\mathcal{E}_2}} & = & \frac{1}{2} tr(\bar{B}_{\bar{3}}
B_{\bar{3}}i\gamma^\mu V_\mu
^T)\nonumber\\
& = & \frac{1}{2f_\pi ^2}\bar{\Lambda}_c\Lambda_c i\gamma^\mu (\pi^0
\partial_\mu \pi^0 + \pi^{-}\partial_\mu \pi^{+} + \pi^{+}\partial_\mu
\pi^{-}) \;.
\end{eqnarray}
According to the heavy quark symmetry, there are four constraint
relations among those six coupling constants of the Lagrangian of
Eq.~(\ref{general-lag}), i.e.,
\begin{eqnarray}
g_6 = 0 \; ,\; g_3 = \frac{\sqrt{3}}{2}g_1\; ,\; g_5 =
-\frac{3}{2}g_1\; ,\; g_4 = -\sqrt{3}g_2\; , \label{couplings}
\end{eqnarray}
which means the number of independent couplings are then reduced to
two. In this work, we employ $g_1$ and $g_2$ for the numerical
evaluation as did in Ref.~\cite{T.M.Yan}.

Here, to get the dominant interaction potential we restrict our
effort only on the $pion$ exchange processes as usual. Notice that
the couplings of $pion$ to spin-$\frac{3}{2}$ and -$\frac{1}{2}$
baryons, and $pion$ to two spin-$\frac{1}{2}$ baryons take a similar
form, in the following we merely present the spin-$\frac{3}{2}$ and
-$\frac{1}{2}$ baryon-$pion$ coupling for illustration, i.e.,
\begin{equation}
\mathcal{L}_1=\frac{g_3}{\sqrt{2}f_\pi}
\bar{\Sigma_c}\!^{0*\mu}\partial_\mu\pi^0\Sigma_c^0 + h.c.\;
,\label{vertex1}
\end{equation}
\begin{equation}
\mathcal{L}_2=-\frac{g_3}{\sqrt{2}f_\pi}
\bar{\Sigma_c}\!^{+*\mu}\partial_\mu\pi^{+}\Sigma_c^{0} + h.c.\;
,\label{vertex2}
\end{equation}
\begin{equation}
\mathcal{L}_3=\frac{g_4}{f_\pi}
\bar{\Sigma_c}\!^{++*\mu}\partial_\mu\pi^{+}\Lambda_c^{+} + h.c.\;
,\label{vertex3}
\end{equation}
\begin{equation}
\mathcal{L}_4=-\frac{g_4}{f_\pi}
\bar{\Sigma_c}\!^{0*\mu}\partial_\mu\pi^{-}\Lambda_c^{+} + h.c.\;
,\label{vertex4}
\end{equation}
\begin{equation}
\mathcal{L}_5=-\frac{g_4}{f_\pi}\bar{\Sigma_c}\!^{+*\mu}
\partial_\mu\pi^{0}\Lambda_c^{+} + h.c.\; .\label{vertex5}
\end{equation}
To get the $pion$ and two spin-$\frac{1}{2}$ baryon couplings one
only needs to replace the $\Sigma_c^{*\mu}$ by $\Sigma_c$, $g_3$ by
$g_1$, $g_4$ by $g_2$, and insert $\gamma^\mu \gamma_5$ in between
the two baryon fields in Eqs.(\ref{vertex1})-(\ref{vertex5}).
\begin{figure}[t,m]
\begin{center}
\scalebox{0.50}{\includegraphics{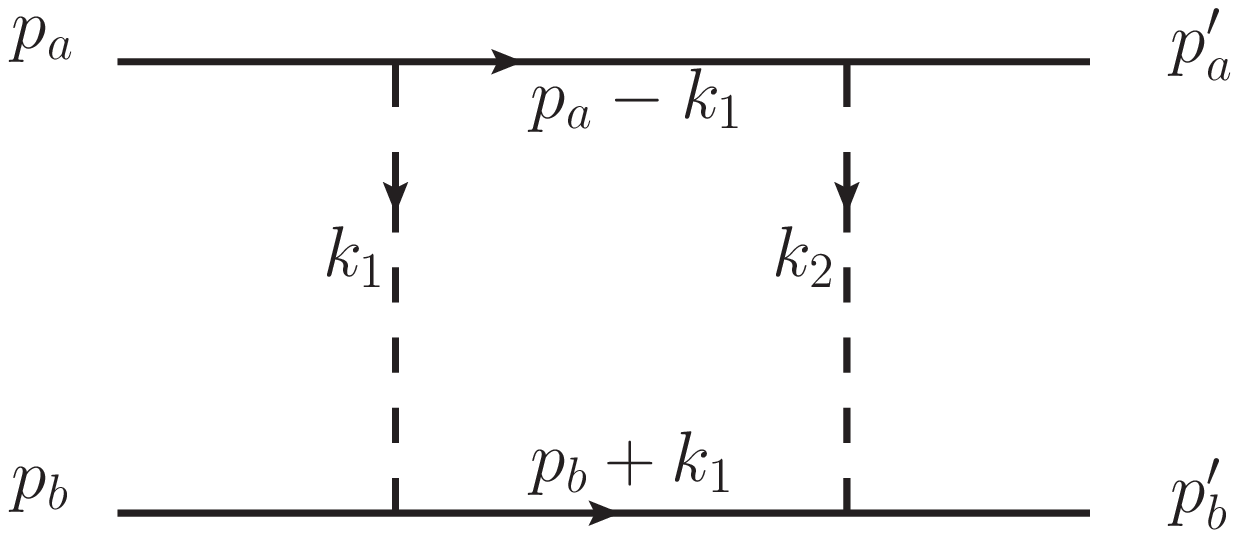}}%
\scalebox{0.50}{\includegraphics{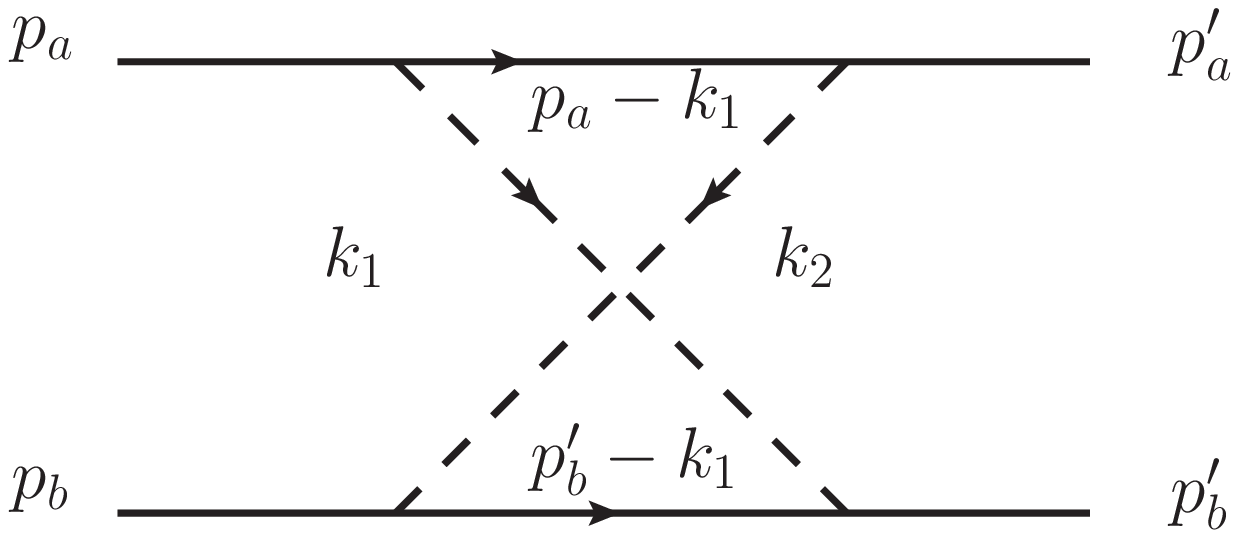}}%
 \caption{Schematic Diagrams which contribute
  to the baryonium potential.}
  \label{fig-bc}
  \end{center}
\end{figure}

\subsection{Baryonium Potential from Two-pion Exchange}

To obtain heavy baryon-baryon interaction potential in configuration
space, we start from writing down the two-body scattering amplitude
in the center-of-mass frame(CMS), i.e. taking $\textbf{p}_a = -
\textbf{p}_b$ and $\textbf{p}_a' = -\textbf{p}_b'$. In CMS the total
and relative four momenta are defined as
\begin{eqnarray}
P & = &(p_a\; +\; p_b)\; =\; (p_a'\; +\; p_b')=(E,\; 0)\; ,\\
p & = &\frac{1}{2}(p_a \;-\; p_b)\; =\; (0,\;\textbf{p})\; ,\\
p'& = &\frac{1}{2}(p_a'\; -\; p_b')\; =\; (0,\; \textbf{p}')\; .
\end{eqnarray}
To perform the calculation, it is convenient to introduce some new
variables as functions of $\textbf{p}$ and $\textbf{p}'$, i.e.,
\begin{eqnarray}
&\mathcal{W}(\textbf{p})& = E_a(\textbf{p})+E_b(\textbf{p})\; ,\\
&\mathcal{W}(\textbf{p}')& = E_a(\textbf{p}')+E_b(\textbf{p}')\; ,\\
&F_E(\textbf{p},\; p_0)& = \frac{1}{2}E+p_0-E(\textbf{p})+i\delta\;
,
\end{eqnarray}
where $\delta$ is an infinitesimal quantity introduced in the
so-called $i\delta$ prescription. Following the same procedure as in
Refs.~\cite{Th.Rijken1,Th.Rijken2}, it is straightforward to write
down the baryon-baryon scattering kernels, shown as box and crossed
diagrams in Figure \ref{fig-bc},
\begin{eqnarray}
K_{box}=&-&\frac{1}{(2\pi)^2}(E-\mathcal{W}
(\textbf{p}'))(E-\mathcal{W}(\textbf{p}))\int
dp_0' dp_0 dk_{20} dk_{10}d^3 \textbf{k}_2d^3 \textbf{k}_1\nonumber\\
& \times& \frac{i}{(2\pi)^4}\delta^4 (p - p'- k_1 - k_2)
\frac{1}{k_2^2 - m^2+i\delta} \frac{1}{F_E(\textbf{p}',p_0')
F_E(-\textbf{p}',-p_0')}\nonumber\\
&\times&\frac{\Gamma_j\Gamma_i\Gamma_i\Gamma_j} {F_E(\textbf{p} -
\textbf{k},p_0-k_{10}) F_E(-\textbf{p}+\textbf{k},-p_0+k_{10})}
\frac{1}{F_E(\textbf{p},p_0) F_E(\textbf{p}, -p_0))}\nonumber\\
&\times&\frac{1} {k_1^2-m^2+i\delta}\; ,\label{kbox}
\end{eqnarray}
\begin{eqnarray}
K_{cross} = & - & \frac{1}{(2\pi)^2}(E-\mathcal{W}
(\textbf{p}'))(E-\mathcal{W}(\textbf{p}))\int
dp_0' dp_0dk_{20}dk_{10}d^3 \textbf{k}_2d^3 \textbf{k}_1\nonumber\\
&\times&\frac{i}{(2\pi)^4} \delta^4(p - p'- k_1 - k_2)
\frac{1}{k_2^2 - m^2 + i\delta}
\frac{1}{F_E(\textbf{p}', p_0') F_E(- \textbf{p}',-p_0')}\nonumber\\
&\times&\frac{\Gamma_j\Gamma_i\Gamma_j\Gamma_i} {F_E(\textbf{p} -
\textbf{k}, p_0-k_{10}) F_E(-\textbf{p}'- \textbf{k},
-p_0' - k_{10})} \frac{1}
{F_E(\textbf{p}, p_0)F_E(-\textbf{p},-p_0)}\nonumber\\
& \times&\frac{1} {k_1^2-m^2 + i\delta}\; . \label{kcross}
\end{eqnarray}
Here, $m$ corresponds to the $pion$ mass and $\Gamma_{i,j}$ are
heavy baryon-$pion$ interaction vertices that can be read out from
the Lagrangian in Eqs.(\ref{vertex1})-(\ref{vertex5}). In case of
spin-$\frac{3}{2}$ intermediate,
\begin{eqnarray}
\Gamma_j\Gamma_i\Gamma_i\Gamma_j & =
&\left(\frac{g_4}{f_\pi}\right)^4\bar{u}(-p)k_2^\mu
u_\mu(p-k_1)\bar{u}_\nu(p-k_1)k_1^\nu
u(p)\nonumber\\
& \times &\bar{v}(p)(-k_1^\alpha) v_\alpha(-p+k_1)\bar{v}_\beta(-p+
k_1)k_2^\beta v(-p)\; ,
\end{eqnarray}
and in case of spin-$\frac{1}{2}$ intermediate
\begin{eqnarray}
\Gamma_j\Gamma_i\Gamma_i\Gamma_j&=& \left(\frac{g_2}{f_\pi}\right)^4
\bar{u}(-p)\gamma_\mu\gamma_5 k_2^\mu u(p-k_1)
\bar{u}(p-k_1)\gamma_\nu\gamma_5 k_1^\nu
u(p)\nonumber\\
& \times &\bar{v}(p)\gamma_\alpha\gamma_5 (-k_1^\alpha) v(-p+k_1)
\bar{v}(-p+k_1)\gamma_\beta\gamma_5 k_2^\beta v(-p)\; .
\end{eqnarray}

Integrating over $p'_0$, $p_0$, $k_{10}$, and $k_{20}$ in
Eq.(\ref{kbox}) one obtains the interaction kernel of box diagram at
order $\mathcal{O}(\frac{1}{M_H})$,
\begin{eqnarray}
K_{box}=&-&\frac{1}{(2\pi)^3}\int\frac{d^3 \textbf{k}_1 d^3
\textbf{k}_2}{4E_{\textbf{k}_1}E_{\textbf{k}_2}}
\frac{\Gamma_j\Gamma_i}
{E_{\textbf{p}-\textbf{k}_1}+E_{\textbf{p}}-W+E_{\textbf{k}_1}}\nonumber\\
&\times&\frac{\Gamma_i\Gamma_j}
{E_\textbf{p}'+E_{\textbf{p}-\textbf{k}_1}-W+E_{\textbf{k}_2}}
\frac{1}{E_{\textbf{p}}+E_{\textbf{p}'}
-W+E_{\textbf{k}_1}+E_{\textbf{k}_2}}\;,
\end{eqnarray}
where $M_H$ represents one of the heavy baryon mass,
$M_{\Lambda_c^+}$, $M_{\Sigma^0_c}$ or $M_{\Sigma_c^*}$;
$E_{\textbf{p}-\textbf{k}_1}=
\sqrt{(\textbf{p}-\textbf{k}_1)^2+M_{\Sigma_c^*}^2}$ is the
intermediate state energy;
$E_{\textbf{k}_1}=\sqrt{\textbf{k}_1^2+m^2}$ and
$E_{\textbf{k}_2}=\sqrt{\textbf{k}_2^2+m^2}$ are two $\it{pion}$s'
energies; and $W = 2 E(\textbf{p})$. With the same procedure, we can
get the interaction kernel of crossed diagram, i.e.
\begin{eqnarray}
K_{cross}=&-&\frac{1}{(2\pi)^3}\int\frac{d^3 \textbf{k}_1 d^3
\textbf{k}_2}{4E_{\textbf{k}_1}E_{\textbf{k}_2}}
\frac{\Gamma_j\Gamma_i} {E_{\textbf{p}-\textbf{k}_1}+
E_{\textbf{p}}-W+E_{\textbf{k}_1}}\nonumber\\
&\times&\frac{\Gamma_j\Gamma_i} {E_\textbf{p}'+E_{\textbf{p}'+
\textbf{k}_1}-W+E_{\textbf{k}_1}} \frac{1}{E_{\textbf{p}}+
E_{\textbf{p}'}-W+E_{\textbf{k}_1}+E_{\textbf{k}_2}}\;
.\label{cross}
\end{eqnarray}

Next, since what we are interested in is the heavy baryons, we can
further implement the non-relativistic reduction on spinors with the
help of vertices given in Eqs.(\ref{vertex1})-(\ref{vertex5}). In
the end, the non-relativistic reduction for $\Lambda_c^ +
\Sigma_c^{+*}\pi^0$ and $\Lambda_c^+\Sigma_c^{+}\pi^0$ couplings
gives
\begin{equation}
i\left(\frac{g_4}{f_\pi}\right)\bar{u}(p_2) u_\mu(p_1)(p_2-p_1)^\mu
= -i\left(\frac{g_4}
{f_\pi}\right)\textbf{S}^{\dag}\cdot\textbf{q}\; ,\label{spin12}
\end{equation}
and
\begin{equation}
i\left(\frac{g_2}{f_\pi}\right)\bar{u}(p_2)\gamma_\mu\gamma_5
u(p_1)(p_2-p_1)^\mu=i\left(\frac{g_2}
{f_\pi}\right)\boldsymbol{\sigma}_1\cdot\textbf{q}\; ,\label{spin32}
\end{equation}
respectively. Here, $\textbf{q}=\textbf{p}_2-\textbf{p}_1$ and
$\textbf{S}^{\dag}$ is the spin-$\frac{1}{2}$ to spin-$\frac{3}{2}$
transition operator.

In the process of deriving $\Lambda_c^+-\bar{\Lambda}_c^+$
potential, the $\Sigma_c^+$ and $\Sigma_c^{+*}$ are taken into
account as intermediate states. Using Eqs.
(\ref{spin12})-(\ref{spin32}) and the explicit forms of spinors
given in the appendix, we can readily obtain the reduction forms for
the $\Sigma_c^+$ intermediate
\begin{eqnarray}
&&\bar{u}(-p)\gamma_\mu\gamma_5 k_2^\mu u(p-k_1)
\bar{u}(p-k_1)\gamma_\nu\gamma_5 k_1^\nu
u(p)\times\nonumber\\
&& \bar{v}(p)\gamma_\alpha\gamma_5 (-k_1^\alpha) v(-p+k_1)
\bar{v}(-p+k_1)\gamma_\beta\gamma_5
k_2^\beta v(-p)\nonumber\\
& = &(\textbf{k}_1\cdot\textbf{k}_2)^2+
(\boldsymbol{\sigma}_1\cdot\textbf{k}_1
\times\textbf{k}_2)(\boldsymbol{\sigma}_2\cdot\textbf{k}_1
\times\textbf{k}_2)\;,
\end{eqnarray}
the $\Sigma_c^{+*}$ intermediate in the box diagram
\begin{eqnarray}
&&\bar{u}(-p)k_2^\mu u_\mu(p-k_1) \bar{u}_\nu(p-k_1)k_1^\nu
u(p)\times\nonumber\\
&&\bar{v}(p)(-k_1^\alpha) v_\alpha(-p+k_1)\bar{v}_\beta(-p+
k_1)k_2^\beta v(-p)\nonumber\\
& = &\frac{4}{9}(\textbf{k}_1\cdot\textbf{k}_2)^2-
\frac{1}{9}(\boldsymbol{\sigma}_1\cdot\textbf{k}_1
\times\textbf{k}_2)(\boldsymbol{\sigma}_2\cdot\textbf{k}_1
\times\textbf{k}_2)\; ,
\end{eqnarray}
and the crossed diagram
\begin{eqnarray}
&&\bar{u}(-p)k_2^\mu u_\mu(p-k_1) \bar{u}_\nu(p-k_1)k_1^\nu
u(p)\times\nonumber\\
&&\bar{v}(p)(-k_1^\alpha) v_\alpha(-p+k_1)\bar{v}_\beta(-p+
k_1)k_2^\beta v(-p)\nonumber\\
& = &\frac{4}{9}(\textbf{k}_1\cdot\textbf{k}_2)^2 +
\frac{1}{9}(\boldsymbol{\sigma}_1\cdot\textbf{k}_1
\times\textbf{k}_2)(\boldsymbol{\sigma}_2\cdot\textbf{k}_1
\times\textbf{k}_2)\; ,
\end{eqnarray}
respectively. Thus, the spinor reduction finally leads to an
operator $\mathcal{O}_1(\textbf{k}_1,\; \textbf{k}_2)$, of which the
variables $\textbf{k}_1$ and $\textbf{k}_2$ can be replaced by
gradient operators $\boldsymbol{\nabla}_1$ and
$\boldsymbol{\nabla}_2$ in configuration space and acting on
$\textbf{r}_1$ and $\textbf{r}_2$, respectively. This operator is
expressed as
\begin{eqnarray}
\mathcal{O}_1(\textbf{k}_1,\; \textbf{k}_2)&=&c_1O_1(\textbf{k}_1,\;
\textbf{k}_2)+c_2O_2(\textbf{k}_1,\; \textbf{k}_2)\nonumber\\
&=&c_1(\textbf{k}_1\cdot\textbf{k}_2)^2+
c_2(\boldsymbol{\sigma}_1\cdot\textbf{k}_1
\times\textbf{k}_2)(\boldsymbol{\sigma}_2\cdot\textbf{k}_1
\times\textbf{k}_2)\; .\label{rdo}
\end{eqnarray}
Here, the decomposition coefficients $c_1$ and $c_2$ are given in
Table 1. The first part of Eq.~(\ref{rdo}) may generate the central
potential and the second part may generate the spin-spin coupling
and the tensor potentials, which are explicitly shown in the
Appendix.
\begin{center}
\vspace{-2mm}
\begin{table}[bht]
\caption{\small The values of coefficients $c_1$ and $c_2$ in the
decomposition of operator $O(\textbf{k}_1,\; \textbf{k}_2)$ in
Eq.~(\ref{rdo}). The left one is for the spin-$\frac{1}{2}$
intermediate state case and the right one is for the
spin-$\frac{3}{2}$ case.} 
\vspace{2mm} \centering
\begin{tabular}{|c c c |}\hline $ $spin-1/2 &$~c_1$&
$~c_2$\tabularnewline\hline\hline
box   & ~1  & ~1\\
cross & ~1  & ~1\\\hline\hline
\end{tabular}
\begin{tabular}{|c c c |}\hline
$ $spin-3/2 &$c_1$& $c_2$\tabularnewline\hline\hline
box   &$ 4/9 $  &$-1/9$\\
cross &$ 4/9$   & $~~1/9$\\\hline\hline
\end{tabular}
\end{table}
\end{center}

To get the leading order central potential, e.g. for
$\Lambda_c$-$\bar{\Lambda}_c$ system, we first expand the energy in
powers of $\frac{1}{M_H}$, but keep only the leading term, like
\begin{eqnarray}
\frac{1}{E_{\textbf{p}-\textbf{k}_1}+
E_{\textbf{p}}-W+E_{\textbf{k}_1}}&\approx&
\frac{1}{M_{\Sigma_c^{*}}+
M_{\Lambda_c}-2M_{\Lambda_c}+E_{\textbf{k}_1}} =
\frac{1}{E_{\textbf{k}_1}+\Delta_1},
\end{eqnarray}
where $\Delta_1=M_{\Sigma_c^{*}}-M_{\Lambda_c}$ represents the mass
splitting. By virtue of the factorization in integrals given in the
Appendix, we can then make a double Fourier transformation, i.e.,
\begin{eqnarray}
V_C^B (r_1,\; r_2)= - \left(\frac{g_4^4} {f_\pi^4}\right)\int\int
\frac{d^3\textbf{k}_1d^3\textbf{k}_2}{(2\pi)^6}
\frac{\mathcal{O}_1(\textbf{k}_1,\textbf{k}_2)
e^{i\textbf{k}_1\textbf{r}_1}e^{i\textbf{k}_2\textbf{r}_2}
f(\textbf{k}_1^2)f(\textbf{k}_2^2)}
{2E_{\textbf{k}_1}E_{\textbf{k}_2}
(E_{\textbf{k}_1}+\Delta_1)(E_{\textbf{k}_2}
+\Delta_1)(E_{\textbf{k}_1}+E_{\textbf{k}_2})}\; , \label{cp0}
\end{eqnarray}
where the superscript $B$ denotes the box diagram and the subscript
$C$ means central potential. Similarly, one can get the central
potential from the crossed diagram contribution
\begin{equation}
V_C^C (r_1,\; r_2)= - \left(\frac{g_4^4} {f_\pi^4}\right)\int\int
\frac{d^3\textbf{k}_1d^3\textbf{k}_2}{(2\pi)^6}
\mathcal{O}_1(\textbf{k}_1,\textbf{k}_2)
e^{i\textbf{k}_1\textbf{r}_1}e^{i\textbf{k}_2\textbf{r}_2}
f(\textbf{k}_1^2)f(\textbf{k}_2^2)\ D\; , \label{cross2}
\end{equation}
where the superscript $C$ denote crossed diagram and the subscript
$C$ means central potential, and
\begin{eqnarray}
D&=& \!\!\!\frac{1}{4 E_{\textbf{k}_1}
E_{\textbf{k}_2}}\left[\left(\frac{1}{(E_{\textbf{k}_1} +
\Delta_1)^2} + \frac{1}{(E_{\textbf{k}_2} + \Delta_1)^2}\right)
\frac{1}{E_{\textbf{k}_1} + E_{\textbf{k}_2}}\right.\nonumber\\
 &+&\!\!\! \left(\frac{1}{(E_{\textbf{k}_1} + \Delta_1)^2} \left. +
\frac{1}{(E_{\textbf{k}_2} + \Delta_1)^2} +
\frac{2}{(E_{\textbf{k}_1} + \Delta_1) (E_{\textbf{k}_2} +
\Delta_1)}\right)\frac{1}{E_{\textbf{k}_1} + E_{\textbf{k}_2} + 2
\Delta_1} \right].
\end{eqnarray}
In order to regulate the potentials we have introduced form factors
at each baryon-pion vertex. The resulting $f(\bf k^2)$ form factors
appearing in Eqs.~(\ref{cp0}) and (\ref{cross2}) will be given in
Section 3.

Taking a similar approach as given in above one can readily get the
central potential in other interaction channels and also the tensor
potential. Notice that although there exists the one-pion exchange
contribution in $\Sigma_c$-$\Sigma_c$ system, due to the $\gamma_\mu
\gamma_5$ nature in interaction vertex, it only contributes to
$\boldsymbol{\sigma}_1\cdot \boldsymbol{\sigma}_2$ term, which is
out of our concern in this work. Here we just focus on the central
potential.

\begin{figure}[t,m]
\begin{center}
\scalebox{0.50}{\includegraphics{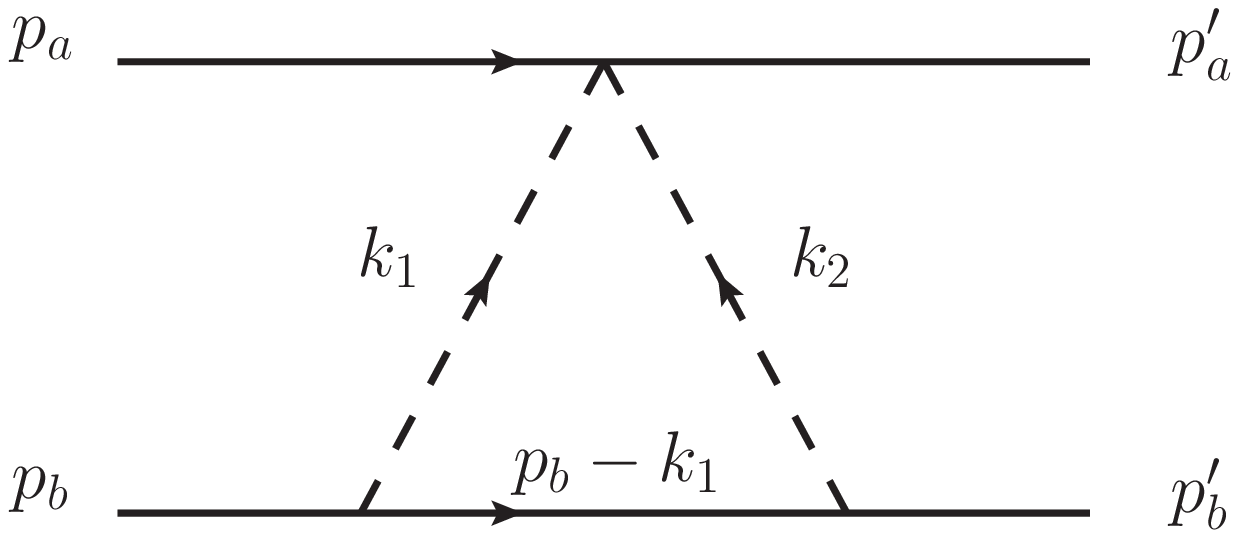}}%
\scalebox{0.50}{\includegraphics{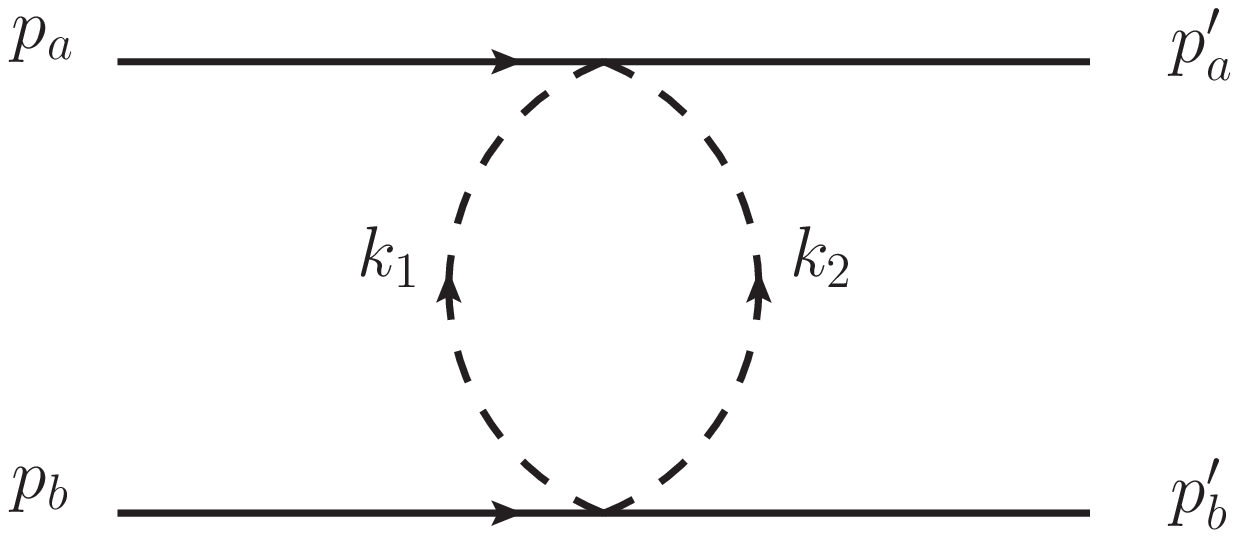}}%
 \caption{The triangle and two-pion loop diagrams.}
 \label{fig-ot}
  \end{center}
\end{figure}
Besides box and crossed diagrams, there are also contributions from
triangle and two-pion loop diagrams as shown in Fig.~\ref{fig-ot}.
As in the box and crossed diagrams, after integrating over energy
component, we get the pion-pair contribution, as shown in the left
diagram of Figure \ref{fig-ot}, as \cite{th_rijken}
\begin{equation}
V_{triangle}(r_1,r_2) = \frac{g_4^2}{2f_\pi^4}\int\int
\frac{d^3\textbf{k}_1d^3\textbf{k}_2}{(2\pi)^6}
\frac{\mathcal{O}_2(\textbf{k}_1,\textbf{k}_2) (E_{\textbf{k}_1} +
E_{\textbf{k}_2})
e^{i\textbf{k}_1\textbf{r}_1}e^{i\textbf{k}_2\textbf{r}_2}
f(\textbf{k}_1^2)f(\textbf{k}_2^2)}{E_{\textbf{k}_1}
E_{\textbf{k}_2}(E_{\textbf{k}_1} + \Delta_1) (E_{\textbf{k}_2} +
\Delta_1)}\;, \label{tri11}
\end{equation}
where the $\mathcal{O}_2(\textbf{k}_1,\textbf{k}_2 )=
(\textbf{k}_1\cdot\textbf{k}_2)$ from spinor reduction can be
replaced in configuration space by the gradient operator
$(\boldsymbol{\nabla}_1\cdot\boldsymbol{\nabla}_2)$. Similarly, the
two-pion loop contribution, as shown in the right diagram of Figure
\ref{fig-ot} reads
\begin{equation}
V_{2\pi-loop}(r_1, r_2) = \frac{1}{16
f_\pi^4}\int\int\frac{d^3\textbf{k}_1d^3\textbf{k}_2}{(2\pi)^6}
e^{i\textbf{k}_1\textbf{r}_1}e^{i\textbf{k}_2\textbf{r}_2}
f(\textbf{k}_1^2)f(\textbf{k}_2^2) A\;. \label{twopair1}
\end{equation}
Here, $A=-\frac{1}{2 E_{\textbf{k}_1}}-\frac{1}{2 E_{\textbf{k}_2}}
+\frac{2}{E_{\textbf{k}_1}+E_{\textbf{k}_2}}~$. Expressing
Eps.~(\ref{tri11}) and (\ref{twopair1}) in the integral
representation of $E_{\textbf{k}_1}$, and making the Fourier
transformation, one can then obtain the corresponding potentials.

\section{Numerical Analysis}

With the central potentials obtained in preceding section, one can
calculate the heavy baryonium spectrum by solving the
Schr\"{o}dinger equation. In our numerical evaluation, the Matlab
based package Matslise \cite{matslise} is employed. The following
inputs from Particle Data Book \cite{PDG} are used in the numerical
calculation:
\begin{equation}
M_{\Lambda_c^+}=2.286 \mathrm{GeV}\; , \;
M_{\Sigma_c^0}=2.454\mathrm{GeV}\; , \;
M_{\Sigma_c^*}=2.518\mathrm{GeV}\; , \; f_\pi=0.132\mathrm{GeV}\;,\;
m = 0.135\mathrm{GeV}\; ,
\end{equation}
and both spin-$\frac{1}{2}$ and -$\frac{3}{2}$ fermion intermediates
are taken into account.

It is obvious that the main uncertainties in the evaluation of heavy
baryonium remain in the couplings of Eq.~(\ref{couplings}). The
magnitudes of the two independent couplings $g_1$ and $g_2$ were
phenomenologically analyzed in Ref.~\cite{T.M.Yan}, and two choices
for them were suggested, i.e.,
\begin{equation}
g_1 = \frac{1}{3}\; , \; g_2=-\sqrt{\frac{2}{3}} \label{para1}
\end{equation}
and
\begin{equation}
g_1 = \frac{1}{3}\times 0.75\; , \; g_2 = -\sqrt{\frac{2}{3}}\times
0.75\; , \label{para2}
\end{equation}
which implies the $g_4$ lies in the scope of 1 to 1.4, similar as
estimated by Ref. \cite{Savage} in the chiral limit.

\subsection{Gaussian form factor case}

The central potential from two-pion exchange box which can be
regularized by widely used Gaussian form factor $f(\textbf{k}^2) =
e^{-\textbf{k}^2/\Lambda^2}$ reads
\begin{eqnarray}
V_{CG}^B(r_1,\; r_2)
&=&-\left(\frac{g_4^4}{f_\pi^4}\right)\left[\frac{1}{\pi}
\int_0^\infty\frac{d\lambda}{\Delta_1^2+\lambda^2}
O_1(\textbf{k}_1,\textbf{k}_2)
F(\lambda,r_1)F(\lambda,r_2)\right.\nonumber\\&&\left.-
\frac{2\Delta_1}{\pi^2}O_1(\textbf{k}_1,\textbf{k}_2)\int_0^\infty
\frac{d\lambda}{\Delta_1^2+\lambda^2}F({\lambda,r_1})
\int_0^\infty\frac{d\lambda}{\Delta_1^2+
\lambda^2}F({\lambda,r_2})\right]\nonumber\\
&=&\sum_i V_{CGi}^B + \cdots\; . \label{cp1}
\end{eqnarray}
Details of the derivation of Eq.~(\ref{cp1}) from Eq.~(\ref{cp0})
can be found in the Appendix. There, the function $F(\lambda,r)$ is
defined by Eq.~(\ref{app72}). And, similarly the central potential
from two-pion exchange crossed diagram gives
\begin{eqnarray}
V_{CG}^C(r_1,\; r_2) &=& -
\left(\frac{g_4^4}{f_\pi^4}\right)\left[\frac{1}{\pi}
\int_0^\infty\frac{d\lambda
(\Delta_1^2-\lambda^2)}{(\Delta_1^2+\lambda^2)^2}
O_1(\textbf{k}_1,\textbf{k}_2)
F(\lambda,r_1)F(\lambda,r_2)\right]\nonumber\\
&=&\sum_i V_{CGi}^C + \cdots\; . \label{crossp1}
\end{eqnarray}
Here, the ellipsis represents the high singular terms in
$r_2\rightarrow r_1=r$ limit, which behave as higher order
corrections to the potential and will not be taken into account in
this work, but will be discussed elsewhere. The central potential of
Eq.~(\ref{cp1}) is obtained in the case of spin-$\frac{3}{2}$
intermediate state, and the explicit forms of $V_{CGi}$ from box
diagram are
\begin{equation}
V_{CG1}^B = -\frac{g_4^4 \Lambda^7}{128 \sqrt{2} \pi^{7/2} f_{\pi}^4
\Delta_1^2} e^{-\frac{\Lambda^2 r^2}{2}}\; ,
\end{equation}
\begin{equation}
V_{CG2}^B = -\frac{g_4^4 \Lambda^5}{16 \sqrt{2} \pi^{7/2} f_{\pi}^4
\Delta_1^2 r^2} e^{-\frac{\Lambda^2 r^2}{2}}\; ,
\end{equation}
\begin{equation}
V_{CG3}^B = \frac{g_4^4 \Lambda^3 m^{5/2} e^{m^2/\Lambda^2}}{32
\sqrt{2} \pi^3 f_{\pi}^4 \Delta_1^2 r^{3/2}} e^{-\frac{\Lambda^2
r^2}{4}-m r}\; ,
\end{equation}
\begin{equation}
V_{CG4}^B = \frac{g_4^4 \Lambda^3 m^{3/2} e^{m^2/\Lambda^2}}{16
\sqrt{2} \pi^3 f_{\pi}^4 \Delta_1^2 r^{5/2}} e^{-\frac{\Lambda^2
r^2}{4}-m r} - \frac{g_4^4 m^{9/2} e^{2 m^2/\Lambda^2}}{128
\pi^{5/2} f_{\pi}^4 \Delta_1^2 r^{5/2}} e^{-2 m r}\; .
\end{equation}
With Gaussian form factors it is seen from Eq.~(\ref{app72}) in the
Appendix that for a given $\Lambda$ the function $F(\lambda,r)$ is
suppressed for large $\lambda$ values, that is the dominant
contribution to potential comes from the small $\lambda$ region. So,
in obtaining the analytic expressions of above potentials and
hereafter, we expand the corresponding functions, as defined in the
Appendix, in $\lambda$ and keep only the leading term. In this
approach, the crossed diagram contributes to the potential the same
as the box diagram at the leading order in $\lambda$ expansion, and
hence is not presented here.

Similarly, we obtain the potentials from triangle and two-pion loop
diagrams, i.e.,
\begin{eqnarray}
V_{CG5}^T &=& \frac{g_4^2 m \Lambda^3}{32 \sqrt{2} \pi^{7/2}f_\pi^4
\Delta_1 r^2}e^{-\frac{\Lambda^2 r^2}{2}} - \frac{g_4^2 m^{5/2}
\Lambda e^{m^2/\Lambda^2}}{16 \sqrt{2} \pi^3f_\pi^4 \Delta_1
r^{5/2}}e^{-\frac{\Lambda^2 r^2}{4}-m r}\nonumber\\
& + & \frac{g_4^2 m^{7/2} e^{2 m^2/\Lambda^2}}{128 \pi^{5/2}f_\pi^4
\Delta_1 r^{5/2}}e^{-2 m r}\;,
\end{eqnarray}
and
\begin{equation}
V_{CG6}^L = - \frac{m^{1/2} \Lambda^3}{32 \sqrt{2} \pi^2 f_\pi^4
r^{3/2}} e^{- \frac{1}{4}\Lambda^2 r^2 -m r }\;.
\end{equation}

To get the central potential for the case of spin-$\frac{1}{2}$
intermediate state, one needs only to make the following replacement
\begin{eqnarray}
g_4\rightarrow g_2\; ,\; \Delta_1\rightarrow \Delta'_1 =
M_{\Sigma_c}-M_{\Lambda_c}
\end{eqnarray}
in Eq.(\ref{cp1}).

\begin{figure}[t,m]
\begin{center}
\scalebox{0.45}{\includegraphics{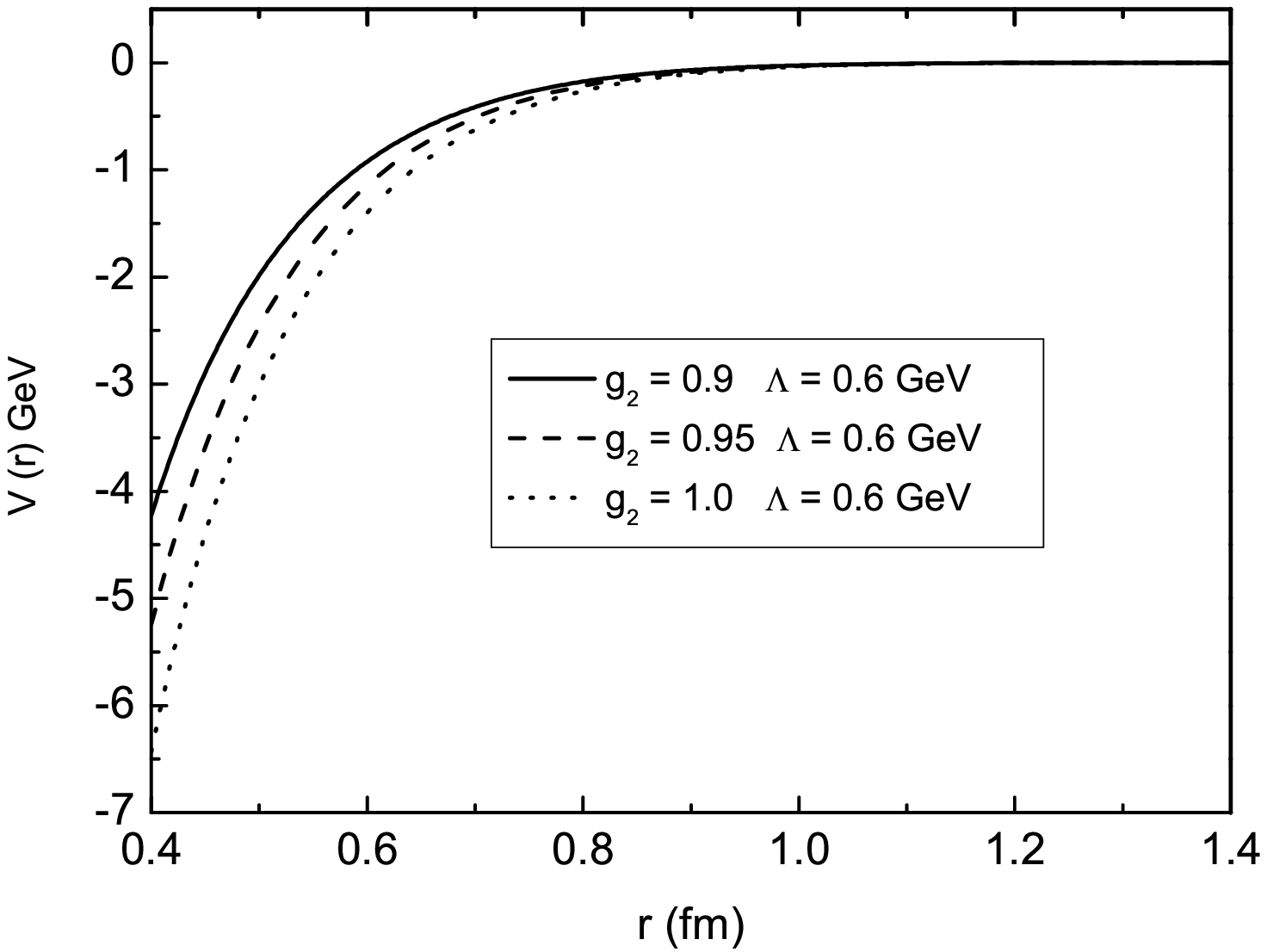}}\hspace{7mm}%
\scalebox{0.45}{\includegraphics{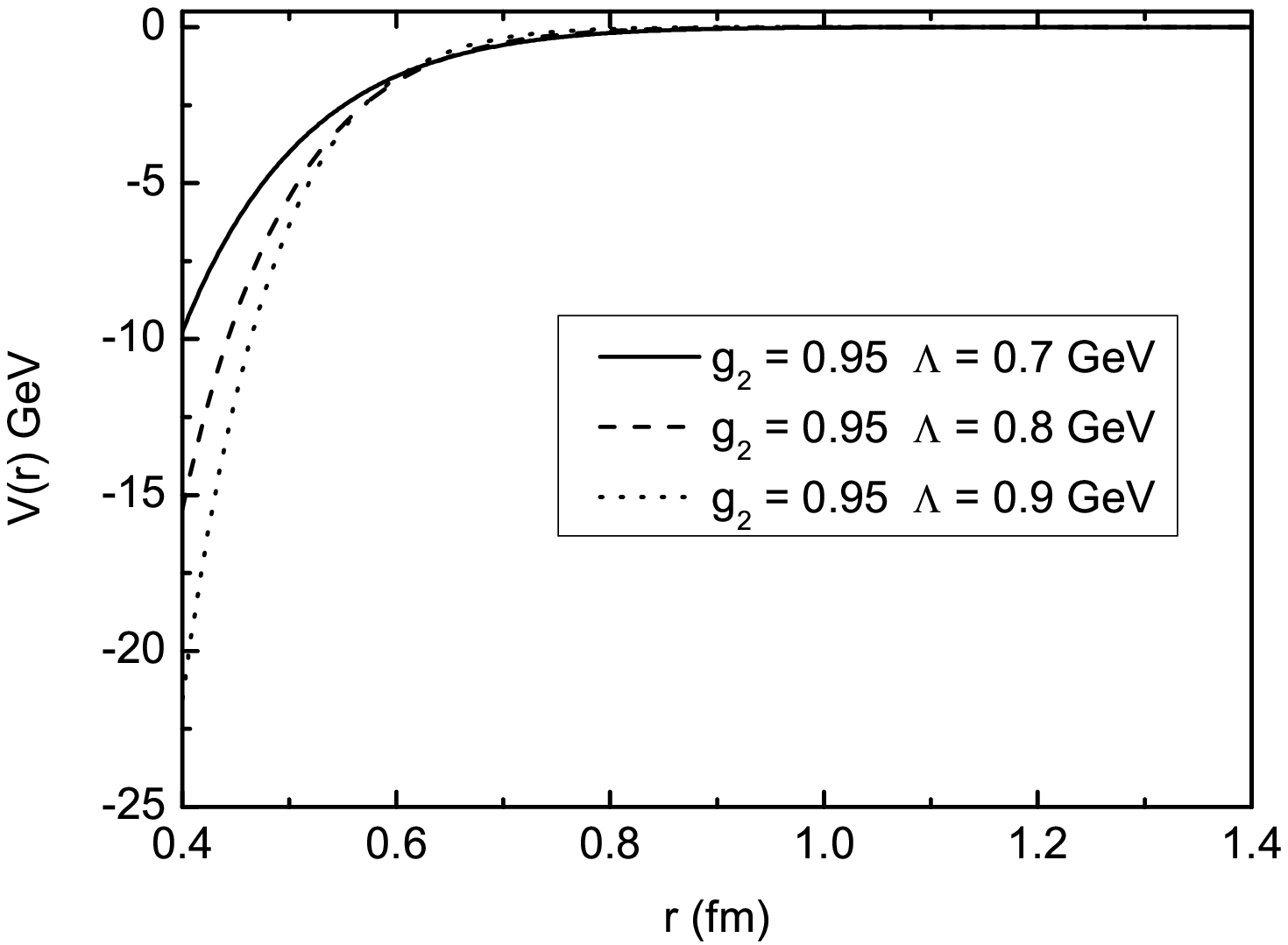}}%
 \caption{The $\Lambda_c$-$\bar{\Lambda}_c$ central potential
  behavior in case of Gaussian form
  factor versus different parameter choices.}
  \label{fig-pgaussian}
  \end{center}
\end{figure}

\begin{figure}[t,m,u]
\begin{center}
\scalebox{0.45}{\includegraphics{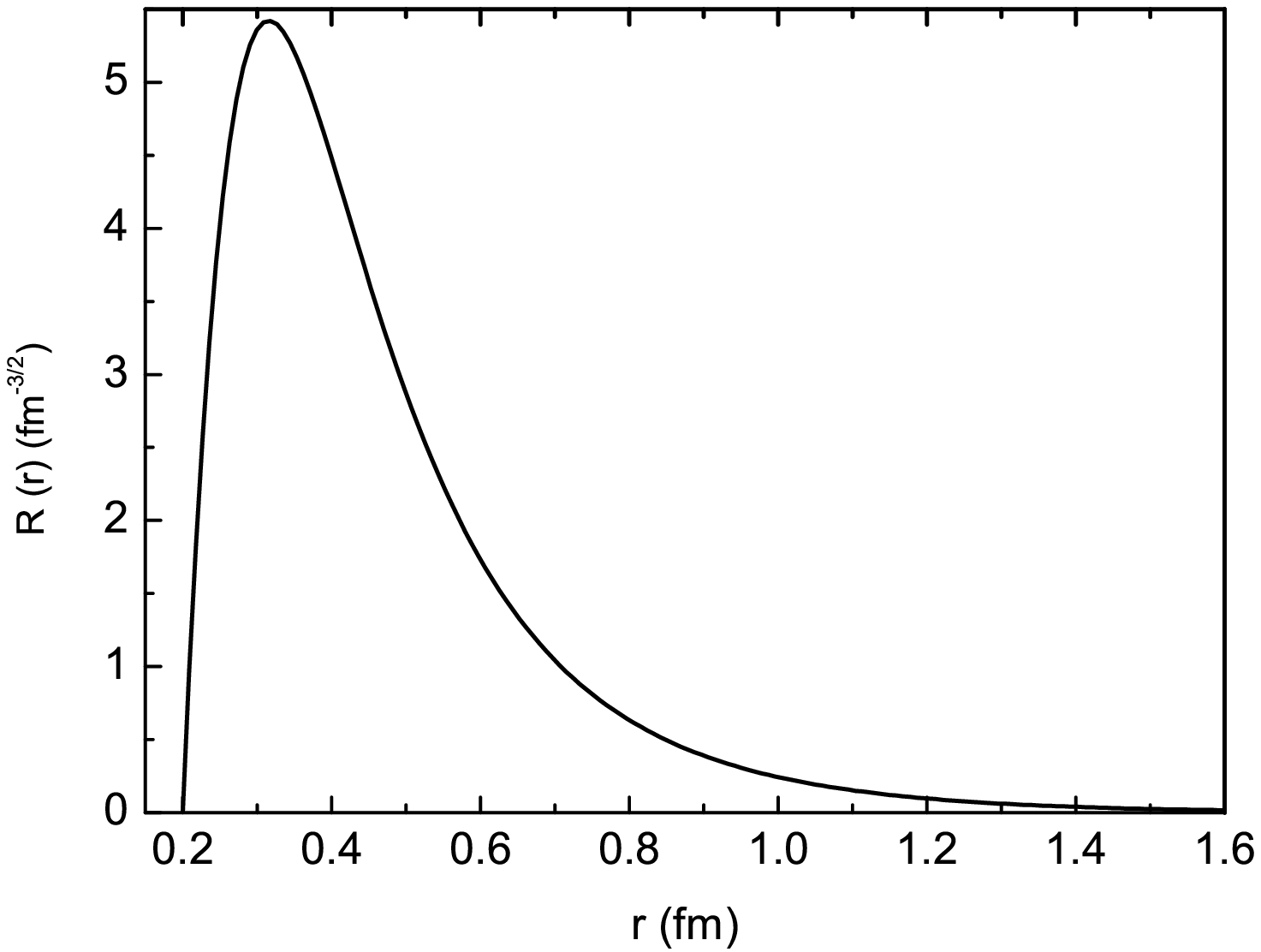}}%
\hspace{10mm}
\scalebox{0.45}{\includegraphics{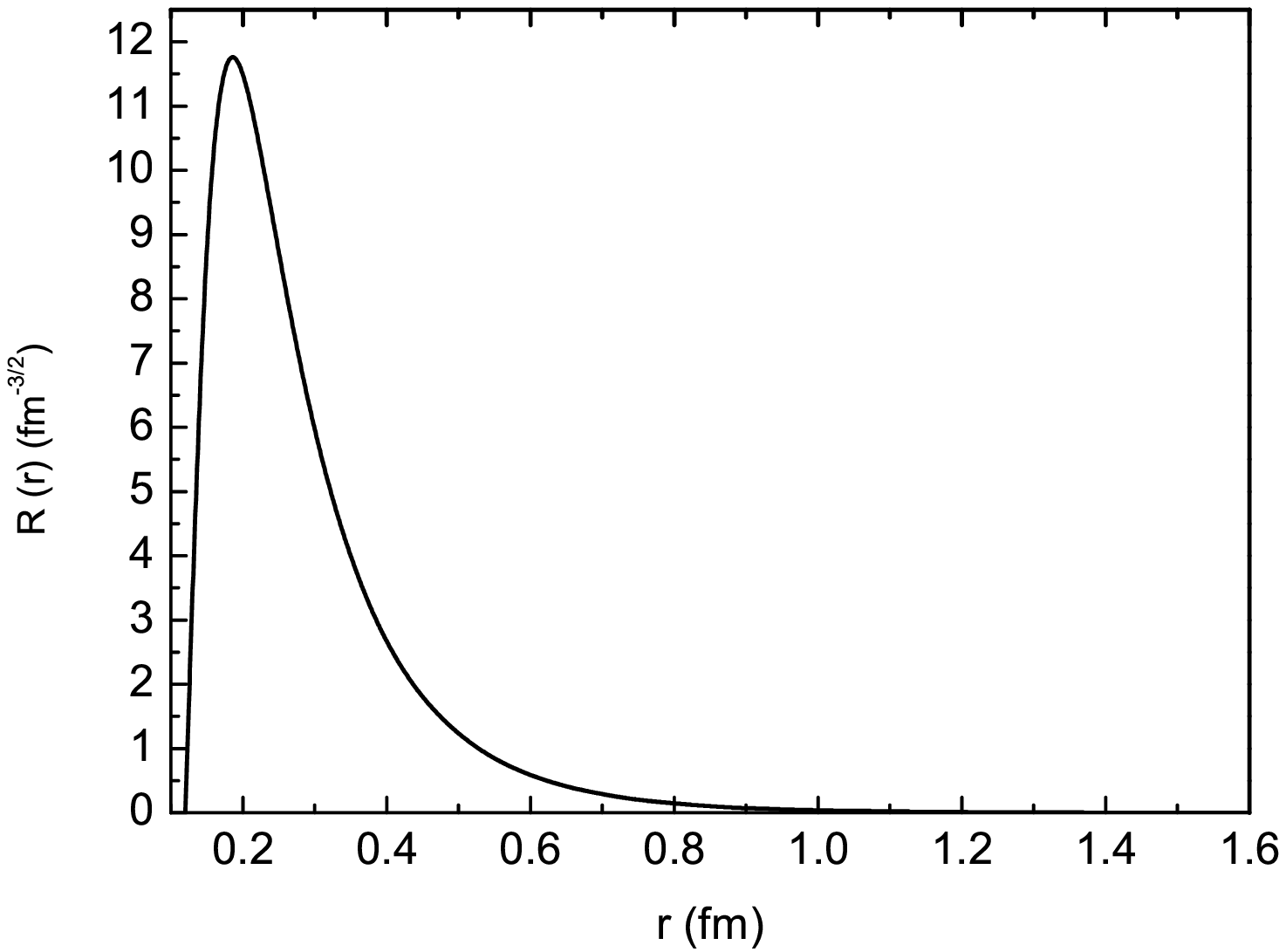}}%
 \caption{Radial wave function of $\Lambda_c$-$
   \bar{\Lambda}_c$ ground state. The left figure is for
   case of Gaussian form factor under the
   condition of $|g_2|=0.95$ and $\Lambda=0.8$, and the
   right one is for the case of monopole form factor with
   $|g_2|=0.9$ and $\Lambda=0.95$.}
  \label{fig-wf}
  \end{center}
\end{figure}

Note that in above asymptotic expressions we keep only those terms
up to order $\frac{1}{r^{5/2}}$, and more singular terms are not
taken into accounted in this work. The dependence of potential with
various parameters are shown in Figure \ref{fig-pgaussian}. The
results indicate that the potential approaches to zero quickly in
long range in every case, while in short range the potential
diverges very much with different parameters, as expected. As a
result, the binding energy heavily depends on input parameters, the
coupling constants and cutoff. One can read from the figure that in
the small coupling situation, the potential becomes too narrow and
shallow to bind two heavy baryons. Table \ref{tb2} presents the
binding energies of $\Lambda_c$-$\bar{\Lambda}_c$ and
$\Sigma_c$-$\bar{\Sigma}_c$ systems with different inputs.
Schematically, the radial wave functions for the ground state of
$\Lambda_c$-$\bar{\Lambda}_c$ system with Gaussian and monopole form
factors are shown in Figure \ref{fig-wf} respectively, while the
wave functions for $\Sigma_c$-$\bar{\Sigma}_c$ system exhibit
similar curves.

\begin{center}
\begin{table}[htb]
\caption{\small Binding energies with different inputs with Gaussian
form factor. The left table is for the $\Lambda_c$-$\bar{\Lambda}_c$
system, and the right one for $\Sigma_c$-$\bar{\Sigma}_c$
system.\vspace{3mm}} \centering
\begin{tabular}{|c c c c |}\hline
$|g_2|$  & $\Lambda(\mathrm{GeV})$& Binding & Baryonium\\
& &energy & mass\tabularnewline\hline\hline
 $<$0.9 &$<$0.6 & No & -\\
0.9  &0.6 &-22 MeV  &  4.550 GeV\\
0.95 &0.6 &-77 MeV  &  4.495 GeV\\
1.0  &0.6 &-168 MeV &  4.404 GeV\\\hline
0.95 &0.7 &-196 MeV &  4.376 GeV\\
0.95 &0.8 &-227 MeV &  4.345  GeV\\
0.95 &0.9 &-588 MeV &  3.984 GeV\\\hline
\end{tabular}
\begin{tabular}{|c  c  c  c|}\hline
$g_1$ & $\Lambda(\mathrm{GeV}) $& Binding & Baryonium\\
& & energy & mass\tabularnewline\hline\hline
$<1.0$ &$<0.8$ & No    & -\\
1.0   &0.8     & -11 MeV &  4.895 GeV\\
1.05  &0.8     &-61 MeV & 4.845 GeV\\
1.1   &0.8     &-145 MeV & 4.761 GeV\\\hline
1.05   &0.85     &-141 MeV & 4.765 GeV\\
1.05   &0.9    &-266 MeV  & 4.640 GeV\\
1.05   &0.95     & -438 MeV & 4.468 GeV\\\hline
\end{tabular}
\label{tb2}
\end{table}
\end{center}

\subsection{Monopole form factor case}

In order to regulate the singularities at the origin in
configuration space, usually people employ three types form factors
in the literature, i.e. the Gaussian, the monopole, and the dipole
form factors \cite{stoks}. For comparison we also calculate the
potential with monopole form factor using the same factorization
technique, and the basic Fourier transformation for monopole form
factor is presented in Appendix for the sake of convenience. Here,
in obtaining the analytic expressions for potentials we also take
the measure of expanding the corresponding functions in parameter
$\lambda$ and keeping only the leading term. Then, what obtained
from the box-diagram contribution reads
\begin{eqnarray}
V_{CM}^B(r)= &-& \frac{g_4^4}{8 \pi^{5/2}f_\pi^4 \Delta^2 r^{5/2}}
\left(\frac{m ^{9/2}}{4} e^{-2 m r} + \frac{\Lambda^4 m ^{1/2}}{4}
e^{-2
\Lambda r}\right) \nonumber\\
&+& \frac{g_4^4 \Lambda^{5/2} m^{5/2}}{8\sqrt{2} \pi^{5/2}f_\pi^4
\sqrt{m+\Lambda}\Delta_1^2 r^{5/2}} e^{-(m + \Lambda) r}\;.
\end{eqnarray}
Contributions from triangle and two-pion loop diagrams are
\begin{eqnarray}
V_{CM}^T (r) &=& \frac{g_4^2 m^{7/2}}{16 \pi^{5/2} f_\pi^4 \Delta_1
r^{5/2}} e^{-2mr} + \frac{g_4^2 m \Lambda^{5/2}}{16 \pi^{5/2}
f_\pi^4 \Delta_1 r^{5/2}} e^{-2\Lambda
r}\nonumber\\
&-& \frac{g_4^2 m^{5/2}\Lambda^{3/2}}{4\sqrt{2} \pi^{5/2} f_\pi^4
 \sqrt{m + \Lambda} \Delta_1 r^{5/2}}  e^{-(m + \Lambda)r}\;
\end{eqnarray}
and
\begin{equation}
V_{CM}^L (r) = -\frac{(\Lambda^2 - m^2) m^{1/2}}{32 \sqrt{2}
\pi^{3/2} f_\pi^4 r^{3/2}}e^{-(m + \Lambda) r} + \frac{(\Lambda^2 -
m^2) \Lambda^{1/2}}{32\sqrt{2} \pi^{3/2} f_\pi^4 r^{3/2}}e^{-2
\Lambda r}\;
\end{equation}
respectively, where superscript $B$, $T$, and $L$ stand for box,
triangle and $2\pi$ loop. Note that since there is no heavy baryon
intermediate state in the $2\pi$ loop process, as shown in the right
graph of Figure 2, the potential range of it appears different.

\begin{figure}[t,m]
\begin{center}
\scalebox{0.45}{\includegraphics{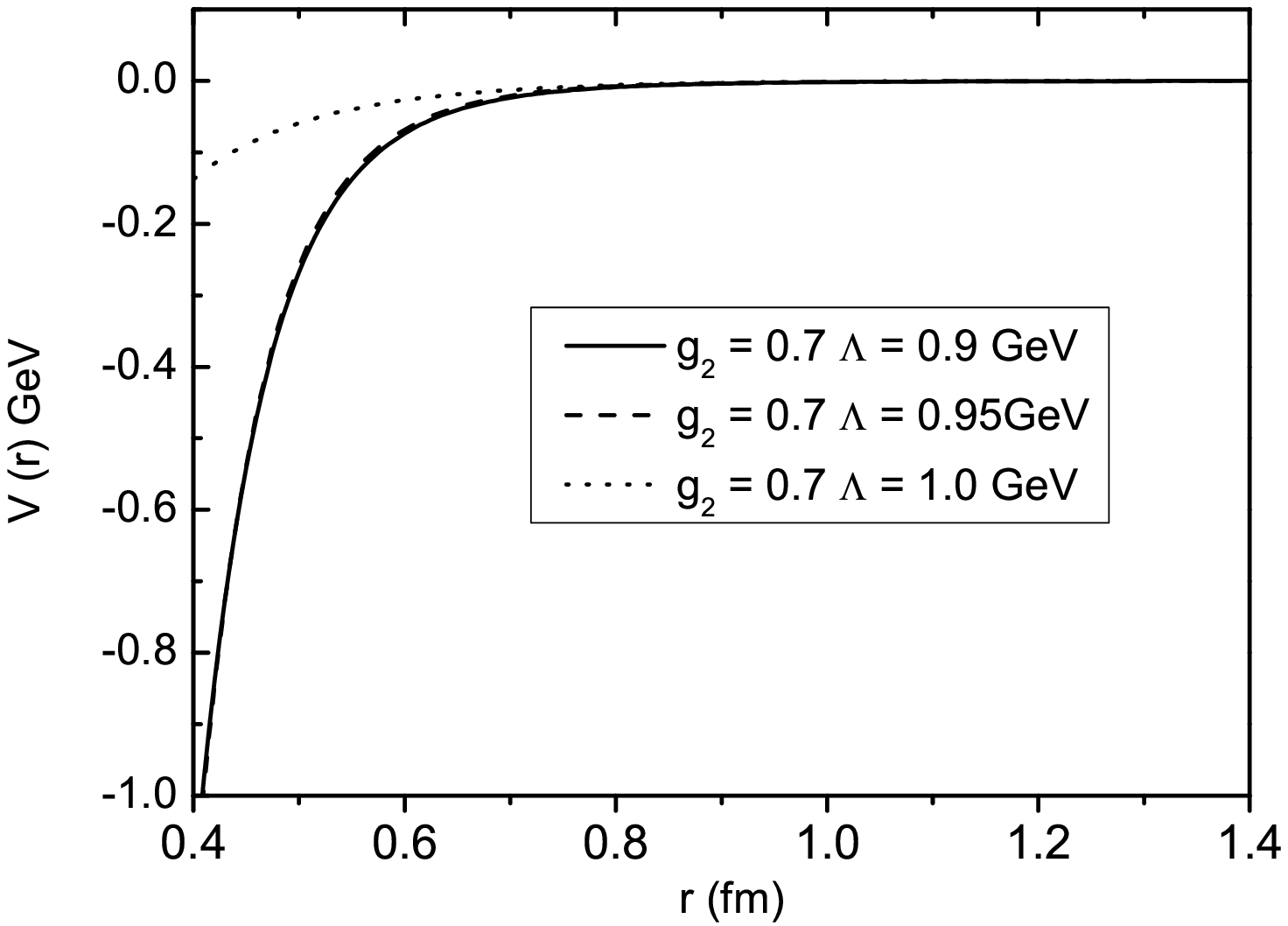}}\hspace{7mm}%
\scalebox{0.45}{\includegraphics{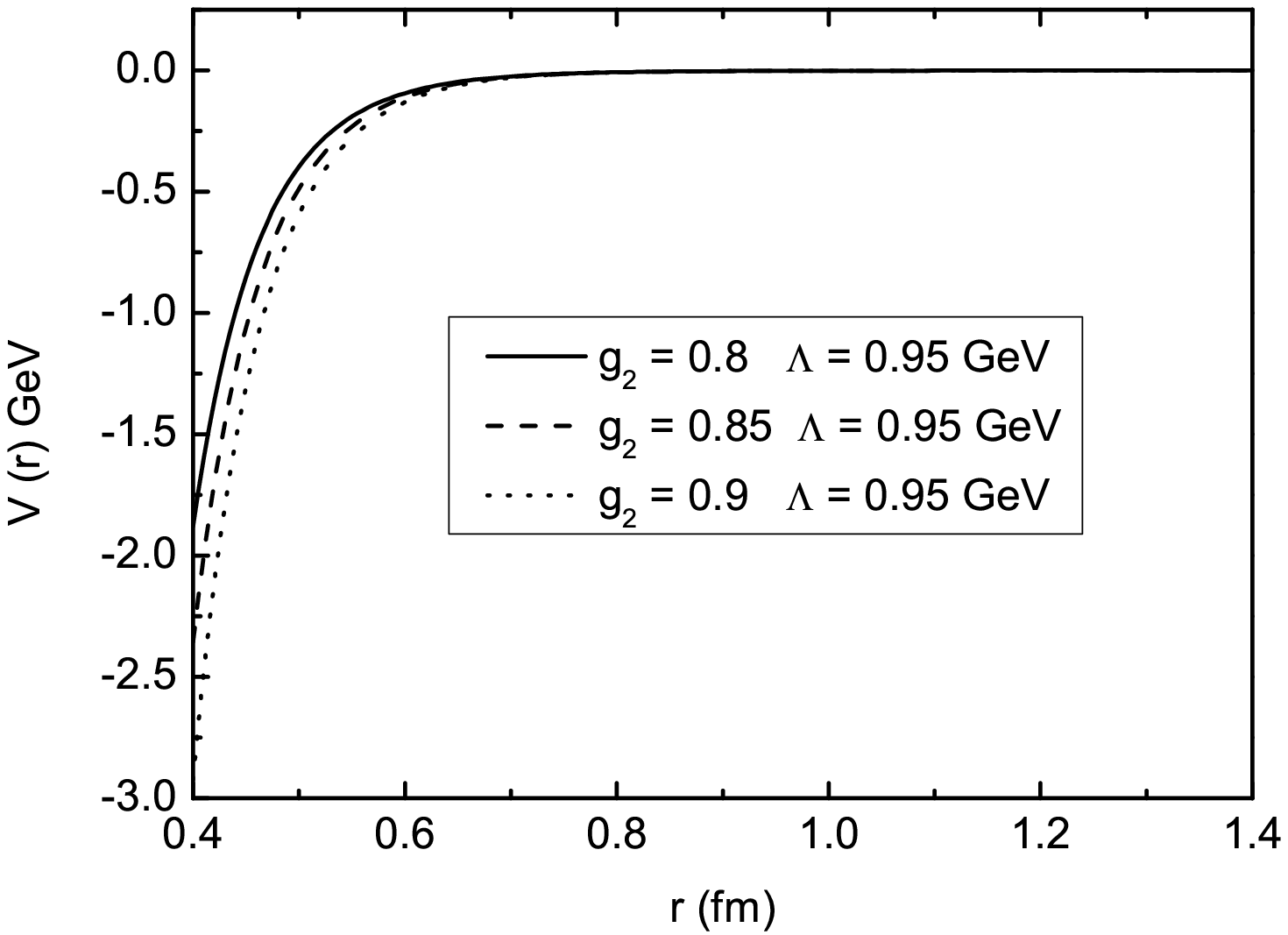}}%
 \caption{The $\Lambda_c$-$\bar{\Lambda}_c$ central potential
  behavior in case of monopole form factor
  versus different choices of inputs.}
  \label{fig-pmono}
  \end{center}
\end{figure}

We find that the structure of potential with monopole form factor is
much simpler than the Gaussian case. The dependence of potential
with various parameters are shown in Fig.\ref{fig-pmono}. From the
figure one can see that in small coupling case the potential change
less, which means the potential tends to be insensitive to the small
coupling, and hence the binding energy. Solving the Schr\"{o}dinger
equation we then obtain eigenvalues for different input parameters,
given in Table \ref{tb3}. From the table, we notice that the binding
energy is sensitive to and changes greatly with the variation of
$g_1$, $|g_2|$ and the cutoff $\Lambda$, the same as the case with
Gaussian form factor. Intuitively, the realistic baryonium can only
accommodate small ones of those parameters.

\begin{center}
\begin{table}[htb]
\caption{\small Binding energies with different inputs with monopole
form factor. The left table is for the $\Lambda_c$-$\bar{\Lambda}_c$
system, and the right one for $\Sigma_c$-$\bar{\Sigma}_c$
system.\vspace{3mm}} \centering
\begin{tabular}{|c c c c |}\hline
$|g_2|$ & $\Lambda(\mathrm{GeV})$& Binding  & Baryonium
\\& & energy & mass\tabularnewline\hline\hline
 $<$0.7 &$<$0.9   &  No & -\\
 0.8 &0.95 &-117 MeV  & 4.455 GeV \\
0.85 &0.95 &-420 MeV  & 4.152 GeV \\
0.9 &0.95 &-521 MeV  & 4.051 GeV\\\hline
0.7 &0.9 &-5 MeV   & 4.567 GeV \\
0.7 &0.95 &-67 MeV  & 4.505 GeV \\
0.7 &1.0 &-252 MeV & 4.320 GeV\\\hline
\end{tabular}
\begin{tabular}{|c   c  c  c|}\hline
$g_1$ & $\Lambda(\mathrm{GeV})$& Binding & Baryonium\\
& & energy & mass\tabularnewline\hline\hline
$<0.9$  &$<0.9$ & No & -\\
0.95  &0.95  &-438 MeV   & 4.468 GeV \\
1.0  &0.95  &-830 MeV   & 4.076  GeV \\
1.05  &0.95  &-1003 MeV   & 3.903 GeV \\\hline
0.9  &0.9  &-40 MeV  & 4.866 GeV \\
0.9  &0.95  &-153 MeV  & 4.753 GeV \\
0.9  &1.0  &-345 MeV  & 4.561 GeV \\\hline
\end{tabular}
\label{tb3}
\end{table}
\end{center}


\subsection{Ground
state of $\Lambda_b$-$\bar\Lambda_b$ baryonium}

\begin{center}
\begin{table}[htb]
\caption{\small Binding energies with the change of parameters for
$\Lambda_b$-$\bar{\Lambda}_b$ system. The left table is for the
Gaussian form factor, and the right one for the monopole form
factor. Here $g_b$ corresponds to $g_2$ in charmed baryonium
sector\vspace{3mm}} \centering
\begin{tabular}{|c c c c |}\hline
$|g_b|$ &$\Lambda(\mathrm{GeV})$& binding &
Baryonium \\
& & energy & mass \tabularnewline\hline\hline
 $<$0.7 &$<$0.7& No & No\\
0.7 &0.75 &-4 MeV & 11.236 GeV \\
0.8 &0.75&-76 MeV &11.164 GeV \\
0.9 &0.75&-294 MeV &10.946  GeV \\\hline
0.8 &0.8&-164 MeV &11.706   GeV \\
0.8 &0.9&-396 MeV & 10.844 GeV \\
0.8 &1.0&-622  MeV &10.618  GeV \\\hline
\end{tabular}
\begin{tabular}{|c  c  c  c |}\hline
$|g_b|$ & $\Lambda(\mathrm{MeV})$& Binding
&Baryonium\\
& & energy & mass\tabularnewline\hline\hline
$<1.0$  &$<0.8$ & No & No\\
1.0  &0.8  &-11  MeV  &11.229   GeV \\
1.05  &0.8  &-56  Mev  &11.184 GeV\\
1.1  &0.8  &-143  MeV  &11.097  GeV \\\hline \hline
1.05  &0.8  &-103  Mev  &11.137  GeV \\
1.05  &0.9  &-164  MeV  &11.076  GeV \\
1.05  &1.0  &-321 MeV  &10.919 GeV\\\hline
\end{tabular}
\label{tb4}
\end{table}
\end{center}

We also estimate the ground state of $\Lambda_b$-$\bar{\Lambda}_b$
baryonium system with Gaussian and monopole form factors. The result
are shown in Table \ref{tb4}, where $g_b$ corresponds to $g_2$ in
charmed baryonium sector. Note that since the dominant decay mode of
$\Sigma_b$ is to $\Lambda_b \pi$, by which we may constrain the
$\Sigma_b \Lambda_b \pi$ coupling from the experiment result, and
this may shed lights on the further investigation on the nature of
possible baryonium.

\section{Summary and Conclusions}

In the framework of heavy baryon chiral perturbation theory we have
studied the heavy baryon-baryon interaction, and obtained the
interaction potential, the central potential, in the case of
two-pion exchange. The Gaussian and monopole types form factors are
employed to regularized the loop integrals in the calculation. As a
leading order analysis, the tensor potential and higher order
contributions in $\frac{1}{M_H}$ expansion are neglected. As
expected, we found that the potential is sensitive to the
baryon-pion couplings and the energy cutoff $\Lambda$ used in the
form factor.

We apply the obtained potential to the Schr\"odinger equation in
attempting to see whether the attraction of two-pion-exchange
potential is large enough to constrain two heavy baryons into a
baryonium. We find it true for a reasonable choice of cutoff
$\Lambda$ and baryon-pion couplings, which is quite different from
the conclusion of a recent work in the study of $D\bar{D}$ potential
through two-pion exchange \cite{qing-xu}. Since usually the cutoff
$\Lambda$ is taken to be less than the nucleon mass, i.e. about 1
GeV in the literature, in our calculation we adopt a similar value
employed in the nucleon-nucleon case. In Ref.~\cite{qing-xu} authors
took a fixed coupling $g=0.59$ and obtained the binding with a large
cutoff. While in our calculation for the baryonium system with
Gaussian form factor, there will be no binding in case $g_1<1.0$ and
$\Lambda<0.8$. The increase of coupling constant will lead to an
even smaller $\Lambda$ for a given binding energy.

Based on our calculation results it is interesting to note that in
case there exists binding in $\Sigma_c$-$\bar{\Sigma}_c$ system,
with both Gaussian and monopole factors, the coupling $g_1$ will be
much bigger than what conjectured in Ref.~\cite{T.M.Yan}. However,
for $\Lambda_c$-$\bar{\Lambda}_c$ system, to form a bound state the
baryon-Goldstone coupling $g_2$ could be similar in magnitude as
what estimated in the literature.

Notice that the potential depends not only on coupling constants and
cutoff $\Lambda$, it also depends on the types of form factors
employed. Our calculation indicates that the Gaussian form factor
and Monopole form factor are similar in regulating the singularities
at origin, and lead to similar results, with only subtle difference,
for both $\Lambda_c$ and $\Lambda_b$ systems. Numerical result tells
that the heavy baryon-baryon potentials are more sensitive to the
coupling constants in the case of Monopole form factor, but more
sensitive to the cutoff $\Lambda$ in the case of Gaussian form
factor. From our calculation it is tempting to conjecture that the
recently observed states $Y(4260)$ and $Y(4360)$, but not $Y(4660)$
\cite{ex}, in charm sector could be a $\Lambda_c$-$\bar{\Lambda}_c$
bound state with reasonable amount of binding energy, which deserves
a further investigation. Our result also tells that the newly
observed ``exotic" state in bottom sector, the $Y_b(10890)$
\cite{K.F.Chen}, could be treated as the
$\Lambda_b$-$\bar{\Lambda}_b$ bound state, whereas with an extremely
large binding energy.

It is worth emphasizing at this point that although our calculation
result favors the existence of heavy baryonium, it is still hard to
make a definite conclusion yet, especially with only the leading
order two-pion-exchange potential. The potential sensitivity on
coupling constants and energy cutoff also looks unusual and asks for
further investigation. To be more closer to the truth, one needs to
go beyond the leading order of accuracy in $\frac{1}{M_H}$
expansion; one should also investigate the potential while two
baryon-like triquark clusters carry colors as proposed in the heavy
baryonium model \cite{Y4260-Qiao,Qiao}; last, but not least, the
unknown and difficult to evaluate annihilation channel effect on the
heavy baryonium potential should also be clarified, especially for
heavy baryon-antibaryon interaction, which nevertheless could be
phenomenologically parameterized so to reproduce known widths of
some observed states.

\vspace{0.3cm}

{\bf Acknowledgments}

This work was supported in part by the National Natural Science
Foundation of China(NSFC) and by the CAS Key Projects KJCX2-yw-N29
and H92A0200S2.

\newpage
\vspace{0.5cm}

{\bf Appendix} \vspace{.3cm}

\appendix{In this Appendix, we present more detailed formulas and
definitions used for the sake of reader's convenience.}

The $\gamma$ matrices take the following convention
\begin{equation}
\gamma^0=\left(
\begin{array}{ll}
 1 & 0 \\
 0 & -1
\end{array}
\right)\; ,\;
\gamma^i=\left(
\begin{array}{ll}
 0 & \sigma^i \\
 -\sigma^i & 0
\end{array}
\right)\; ,\;
\gamma_5=\left(
\begin{array}{ll}
 0 & 1 \\
 1 & 0
\end{array}
\right)\; .
\end{equation}
And the Dirac spinors for $\Sigma_c$ read as
\begin{eqnarray} u(p) =
\sqrt{\frac{E+M_\Sigma}{2M_\Sigma}}\left(\begin{array}{l}
 \chi_a   \\
 \frac{\boldsymbol{\sigma} \cdot \textbf{p}}{E + M_\Sigma}\chi_a
\end{array}\right)\;,
\end{eqnarray}
where $\chi_a$ is two-component Pauli spinor, and
\begin{eqnarray} v(p) =
\sqrt{\frac{E+M_\Sigma}{2M_\Sigma}}\left(\begin{array}{l}
\frac{\boldsymbol{\sigma} \cdot \textbf{p}}{E + M_\Sigma}\eta_a   \\
 \eta_a
\end{array}\right)\;,
\end{eqnarray}
where $\eta_a = -i\sigma^2\chi_a^*$, and $a = 1,2$.
Spin-$\frac{3}{2}$ field for $\Sigma^{+*}$ is described by
Rarita-Schwinger spinor $u^\mu(p\;,\sigma)$, which can be
constructed by spin-$1$ vector and spin-$\frac{1}{2}$ field
\cite{th_rijken2}, that is
\begin{equation}
u^\mu = \sqrt{\frac{E +
M_{\Sigma^{+*}}}{2M_{\Sigma^{+*}}}}L^{(1)}(p)^\mu_\nu
\left(\begin{array}{l}
 1   \\
 \frac{\boldsymbol{\sigma} \cdot \textbf{p}}{E + M_{\Sigma^{+*}}}
\end{array}\right)S^{\dagger\nu}\psi(\sigma)\;,
\end{equation}
where $\psi(\sigma)$ is four-component Pauli spinor of a
spin-$\frac{3}{2}$ particle, and $L^{(1)}(p)^\mu_\nu$ is the boost
operator for spin-$1$ particle,
\begin{equation}
L^{(1)}(p)^\mu_\nu = \left(\begin{array}{ll}
 \frac{E}{M_{\Sigma^{+*}}} & \hspace{15mm} \frac{p_j}{M_{\Sigma^{+*}}} \\
 \frac{p_i}{M_{\Sigma^{+*}}} & \delta^i_j -
 \frac{p^ip_j}{M_{\Sigma^{+*}} (E + M_{\Sigma^{+*}})}
\end{array}
\right)\;,
\end{equation}
where $i,j$ are indices of the space components of momentum $p$. The
positive- and negative-energy projection operators for
spin-$\frac{1}{2}$ baryon are
\begin{eqnarray}
[\Lambda^+(p)]_{\alpha\beta}=\sum_{{\pm}s}
 u_\alpha(p,s)\overline{u}_\beta(p,s)=
 \left(\frac{p\!\!\!/+
 M_{\Sigma_c}}{2M_{\Sigma_c}}\right)_{\alpha\beta}\;
\end{eqnarray}
and
\begin{eqnarray}
[\Lambda^-(p)]_{\alpha\beta}=-\sum_{{\pm}s}
v_\alpha(p,s)\overline{v}_\beta(p,s)=
\left(\frac{-p\!\!\!/+
M_{\Sigma_c}}{2M_{\Sigma_c}}\right)_{\alpha\beta}\;
,
\end{eqnarray}
respectively.

The positive- and negative-energy projection operators for
spin-$\frac{3}{2}$ baryon are
\begin{eqnarray}
\left[\Lambda^+_{\mu\nu}(p)\right]_{\alpha\beta} &&=
\sum_{{\pm}s}u_{\mu,\; \alpha}(p,s)
\overline{u}_{\nu,\; \beta}(p,s)\nonumber\\
&&=[\frac{p\!\!\!/ + M_{\Sigma_c^*}}{2
M_{\Sigma_c^*}}]_{\alpha\beta}
\left(g_{\mu\nu}-\frac{\gamma_\mu\gamma_\nu}{3} - \frac{2p_\mu
p_\nu}{3M_{\Sigma_c^*}^2} + \frac{p_\mu
\gamma_\nu-p_\nu\gamma_\mu}{3M_{\Sigma_c^*}}\right)\; ,
\end{eqnarray}
and
\begin{eqnarray}
\left[\Lambda^-_{\mu\nu}(p)\right]_{\alpha\beta} &&=-
\sum_{{\pm}s}v_{\mu,\; \alpha}(p,s)
\overline{v}_{\nu,\; \beta}(p,s)\nonumber\\
&&=[\frac{-p\!\!\!/ + M_{\Sigma_c^*}}{2
M_{\Sigma_c^*}}]_{\alpha\beta}
\left(g_{\mu\nu}-\frac{\gamma_\mu\gamma_\nu}{3} - \frac{2p_\mu
p_\nu}{3M_{\Sigma_c^*}^2} + \frac{p_\mu
\gamma_\nu-p_\nu\gamma_\mu}{3M_{\Sigma_c^*}}\right)\; ,
\end{eqnarray}
respectively. Here, $\mu$ and
$\nu$ are Lorentz indices; $\alpha$
and $\beta$ are Dirac spinor indices.

The basic Fourier transformation
with Gaussian form factor reads
\begin{eqnarray}
I_2(m,\; r)&=&\int_{-\infty}^{\infty}
\frac{d^3\textbf{k}}{(2\pi)^3}\frac{e^{i\textbf{kr}}
e^{-\textbf{k}^2/\Lambda^2}}{\textbf{k}^2+m^2}\nonumber\\
&=&\frac{1}{8\pi
r}e^{m^2/\Lambda^2}\left[e^{-mr}erfc\left(-\frac{\Lambda
r}{2}+\frac{m}{\Lambda}\right)-e^{mr}erfc\left(\frac{\Lambda
r}{2}+\frac{m}{\Lambda}\right)\right]\; ,
\end{eqnarray}
and hence
\begin{eqnarray}
F(\lambda,\; r)= \int\frac{d^3\textbf{k}}{(2\pi)^3}
\frac{e^{i\textbf{kr}}e^{-\textbf{k}^2/\Lambda^2}}
{\textbf{k}^2+m^2+\lambda^2} = I_2(\sqrt{m^2+\lambda^2},\; r)
e^{-\lambda^2/\Lambda^2}\; . \label{app72}
\end{eqnarray}

$erfc(x)$ is complementary error function, which is defined as
\begin{equation}
erfc(x) = \frac{2}{\sqrt{\pi}}\int_x^\infty e^{-t^2} dt\; .
\end{equation}

The factorization in double Fourier transformation goes like
\begin{eqnarray}
H_{11}&=&\int\int\frac{d^3\textbf{k}_1d^3\textbf{k}_2}
{(2\pi)^6}\frac{e^{i\textbf{k}_1
\textbf{r}_1}e^{i\textbf{k}_2\textbf{r}_2}
f(\textbf{k}_1^2)f(\textbf{k}_2^2)}{\omega_1\omega_2(\omega_1+a)
(\omega_2+a)(\omega_1+\omega_2)}\nonumber\\
&=&\int\int\frac{d^3\textbf{k}_1d^3\textbf{k}_2}
{(2\pi)^6}\frac{1}{a^2}[\frac{2}{\pi}
\int_0^\infty\frac{e^{i\textbf{k}_1\textbf{r}_1}
e^{i\textbf{k}_2\textbf{r}_2}
f(\textbf{k}_1^2)f(\textbf{k}_2^2)d\lambda}
{(\omega_1^2+\lambda^2)(\omega_2^2+\lambda^2)}\\
&-&\frac{2}{\pi}\int_0^\infty\frac{e^{i\textbf{k}_1\textbf{r}_1}
e^{i\textbf{k}_2\textbf{r}_2}
f(\textbf{k}_1^2)f(\textbf{k}_2^2)\lambda^2d\lambda} {(a^2 +
\lambda^2)(\omega_1^2 + \lambda^2)(\omega_2^2 +
\lambda^2)}]-\frac{1}{a}G_{11}(\lambda,\; r_1)
G_{11}(\lambda,\; r_2)\nonumber\\
&=&\frac{2}{\pi}\int_0^\infty\frac{d\lambda}
{a^2+\lambda^2}F(\lambda,\; r_1)F(\lambda,\; r_2)-
\frac{1}{a}G_{11}(\lambda,\; r_1)G_{11}(\lambda,\; r_2)\; .
\end{eqnarray}
Here,
\begin{eqnarray}
G_{11}& = &\int\frac{d^3\textbf{k}_1}
{(2\pi)^3}\frac{e^{i\textbf{k}_1\textbf{r}}
e^{-\textbf{k}_1^2/\Lambda^2}}{\omega_1(\omega_1+a)}
=\int\frac{d^3\textbf{k}_1}{(2\pi)^3}\frac{2a}{\pi}
\int_0^\infty\frac{e^{i\textbf{k}_1\textbf{r}}
e^{-\textbf{k}_1^2/\Lambda^2}d\lambda}{(a^2+
\lambda^2)(\omega_1^2+\lambda^2)}\nonumber\\
&=&\frac{2a}{\pi}\int_0^\infty
\frac{d\lambda}{(a^2+\lambda^2)}F(\lambda,\; r)\; ,
\end{eqnarray}
and for simplicity we define $\omega_1=\sqrt{\textbf{k}_1^2+m^2}$
and $\omega_2=\sqrt{\textbf{k}_2^2+m^2}$\; .

In the case of the monopole form factor, i.e. $f(\textbf{k}^2) =
\frac{\Lambda^2-m^2}{\Lambda^2+\textbf{k}^2}$, the corresponding
function to $F(\lambda ,\;r)$ reads
\begin{eqnarray}
R(\lambda,\;r) &=&  \int \frac{d^3\textbf{k}}{(2\pi)^3}
\frac{e^{i\textbf{kr}}}{\textbf{k}^2 + m^2 + \lambda^2}
 \frac{\Lambda^2-m^2}{\Lambda^2+\textbf{k}^2 + \lambda^2}\nonumber\\
&=&\frac{1}{4\pi r}\left (e^{- r \sqrt{m^2 + \lambda^2}}-e^{- r
\sqrt{\Lambda^2 + \lambda^2}} \right)\; .
\end{eqnarray}

Operator $O_1(\textbf{k}_1,\; \textbf{k}_2)$ contains two parts. The
first part of $O_1(\textbf{k}_1,\;\textbf{k}_2)$ while acting on
functions in configuration space goes like
\begin{eqnarray}
O_1(\textbf{k}_1,\; \textbf{k}_2)F(\lambda,\; r_1) F(\lambda,\;
r_2)&=&(\textbf{k}_1 \cdot \textbf{k}_2)^2F
(\lambda,\; r_1)F(\lambda,\; r_2)\nonumber\\
&=&(\nabla_{1i}\nabla_{1j})F(\lambda,\; r_1)
(\nabla_{2i}\nabla_{2j})F(\lambda,\; r_2)\nonumber\\
&=&\frac{2}{r^2}F'(\lambda,\;
r)F'(\lambda,\;r)+F''(\lambda,\;r)F''(\lambda,\;r) \; ,
\end{eqnarray}
where
\begin{equation}
\nabla_i\nabla_j=\left(\delta_{ij}-\frac{x_ix_j}{r^2}\right)
\left(\frac{1}{r}\frac{d}{dr}\right)+\frac{x_ix_j}{r^2}
\left(\frac{d^2}{dr^2}\right)\;,
\end{equation}
and the limit $r_2\rightarrow r_1=r$ is taken. The second part of
$O_2(\textbf{k}_1,\;\textbf{k}_2)$ while acting on functions in
configuration space goes like
\begin{eqnarray}
O_2(\textbf{k}_1,\;\textbf{k}_2) F(\lambda,\; r_1) F(\lambda,\; r_2)
& = & (\boldsymbol{\sigma}_1\cdot\textbf{k}_1
\times\textbf{k}_2)(\boldsymbol{\sigma}_2\cdot\textbf{k}_1
\times\textbf{k}_2) F(\lambda,\;r_1) F(\lambda,\;r_2)\nonumber\\
& = &\sigma_{1i} \sigma_{2j}\varepsilon_{ikl} \varepsilon_{jmn}
(\nabla_{1k}\nabla_{1m})F(\lambda,\; r_1)
(\nabla_{2l}\nabla_{2n})F(\lambda,\; r_2)\nonumber\\
& = & \sigma_{1i} \sigma_{2j} (\delta_{ij} \delta_{km} \delta_{ln} +
\delta_{im} \delta_{kn} \delta_{lj} + \delta_{in} \delta_{lm}
\delta_{kj}\nonumber\\
&&- \delta_{lj} \delta_{km} \delta_{in} - \delta_{lm} \delta_{kn}
\delta_{ij} - \delta_{ln} \delta_{im} \delta_{kj})\times\nonumber\\
&&(\nabla_{1k}\nabla_{1m})F(\lambda,\; r_1)
(\nabla_{2l}\nabla_{2n})F(\lambda,\; r_2)\nonumber\\
&=& \frac{2}{3}\left[\frac{1}{r^2} F'(\lambda,\; r) F'(\lambda,\;r)
+ \frac{2}{r}F'(\lambda,\; r) F''(\lambda,\; r)\right]
(\boldsymbol{\sigma}_1\cdot
\boldsymbol{\sigma}_2)\nonumber\\
&& + \frac{2}{3}\left(\frac{F'(\lambda,\;
r)}{r}-F''(\lambda,\;r)\right)\frac{1}{r}F'(\lambda,\;r) S_{12}\;,
\end{eqnarray}
where $\boldsymbol{\sigma}_1\cdot \boldsymbol{\sigma}_2$ gives
spin-spin potential and $S_{12} = \frac{3
(\boldsymbol{\sigma}_1\cdot \textbf{r})(\boldsymbol{\sigma}_2\cdot
\textbf{r})}{r^2} - \boldsymbol{\sigma}_1\cdot
\boldsymbol{\sigma}_2$ gives the tensor potential.

\end{document}